# Impact of micro-credit on the livelihoods of clients: A study on Sunamganj District

**DR MD NAZRUL ISLAM**

**Professor,**

**Department of Business Administration**

**Shahjalal University of Science and Technology**
**Sylhet-3144, Bangladesh**

**Funded by University Research Centre**
Shahjalal University of Science and Technology

# Abstract

The objective of this paper is to assess the impact of micro credit on the livelihoods of the clients in the haor area of Sunamganj district, Sylhet, Bangladesh. The major findings of the study are that (i) 66.2% respondents of borrowers and 98.7 non-borrowers are head of the family and an average 76.6% and among the borrowers 32% is husband/wife while 1.3% of non-borrowers and on average 22.2. In terms of sex 64.7% of borrowers and 92.5% of non-borrowers are male while 35.3% of borrowers and 7.5% of non-borrowers are female. The impact of micro-credit in terms of formal and informal credit receiving households based on DID method showed that total income; total expenditure and investment have been increased 13.57%, 10.39% and 26.17%. All the elements of total income have been increased except debt which has been decreased by 2.39%. But the decrease in debt is the good sign of positive impact of debt. Consumption of food has been increased but non-food has been decreased. All the elements of investment have been increased except some factors. The savings has been decreased due excess increase in investment. The study suggested that for breaking vicious cycle of poverty by micro-credit the duration of loans should be at least five year and the volume of loans must be minimum 500,000 and repayment should at not be less than monthly. The rate of interest should not be more than 5%.

# Summary

The *haor* area is a wetland ecosystem located in the north eastern part of Bangladesh which physically is a bowl or saucer shaped shallow depression, also known as a back swamp. Despite the economic importance of the *haors*, people in the region are poorer than in any other part of the country due to geographical remoteness from the land. In the *haor* areas more than 28% of the total population lives below the Lower Poverty Line (LPL). Geographical remoteness is a constraint of development policy in the developing country and isolation from the growth center is also embarrassing the development plan. Sunamganj district is the mostly *haor*-based district in Bangladesh and purposively this *haor*-based district has been chosen for the study. In Sunamganj district there are 11 upazilas and three of them have been selected for the study as they are known haor-based area nationally and internationally. The three upazilas are: Bishwamvarpur upazila, Tahirpur upazila and Dharmapasha upazila. The core objective of project is to assess the impact of micro credit on the livelihoods of the clients in the *haor* area of Sunamganj district. The specific objectives are to (i) identify the nature of microcredit services available in the *haor* area of Sunamganj district; (ii) measure the productive outcome of microcredit on the livelihoods of the clients; (iii) identify the problems in microcredit facing by the clients in the utilization of microcredit services; and (iv) draw implications for design and implementation of micro-credit programs in the *haor* areas of Bangladesh specially for Sunamganj district. In measuring the impact of microcredit on clients, at first the clients have been divided into two groups on the basis of microcredit inclusion. The first group was included those clients who are microcredit member for at least three years and the second group of respondents is consisted of people who are not members of any microcredit program. Random sampling has been used to select the respondents and from the first group (borrowers) 156 micro credit borrowers (members) from each upazila and the total sample size is 468 (156×3) micro credit borrowers have been selected for the interview process from the three upazilas but in practice we have received 485 micro credit borrowers from survey. We have selected 78 non-borrowers (control group) from each upazila randomly in our interview process and total sample of control group is 78×3 = 234. But in practice we have received 228. Thus, total sample size stood at 713(485+228). A semi-structured questionnaire has been prepared for collecting data from the two groups of respondents as stated: the data from the respondents' *i.e.*, primary data has been collected through field survey by the data collector with the semi structured questionnaire. The secondary data has been collected from the related published and unpublished books, theses, reports and articles. The Statistical Package for Social Sciences (SPSS) software has been used for the statistical analyses. Mainly two kinds of analyses have been made from the data base. Firstly the descriptive statistics include the mean, standard deviation and percentage have been determined of the variables viz. the volume and duration of credit, respondent's annual income, annual expenditure, annual savings, knowledge about micro credit, expenditure or investment pattern of micro credit, purpose of loan causes of non- payment of loan and attitude of borrower's about micro-credit possession assets, possession of productive assets, savings, net worth, expenditure, financial capacity and demographic profile of all two groups Secondly - Difference in differences (DID is used to measure the impact of micro credit on poverty alleviation. Under this method first we calculated the effect of a treatment i.e., an explanatory variable on an outcome (i.e., a response variable or dependent variable) by comparing the average change over time in the outcome variable for the treatment group to the average change over time for the control group. Factor analysis with Principal Component Analysis (PCA) has been used to identify the significant causes and factors for relevant cases on poverty and micro-credit. Finally, two Focus Group Discussions (FGD) were conducted in each Up-zilla participating different representatives of different group. The major findings of the study are that (i) 66.2% respondents of borrowers and 98.7 non-borrowers are head of the family and an average 76.6% and among the borrowers 32% is husband/wife while 1.3% of non-borrowers and on average 22.2. In terms of sex 64.7% of borrowers and 92.5% of non-borrowers are male while 35.3% of borrowers and 7.5% of non-borrowers are female. The age category of respondents showed that 62.5% of borrowers and 66.7% of non–borrowers are 31-50 and on average l 63.8%. The marital status of the respondents is that 92.2% of borrowers and 92.5% of non–borrowers are married and on average l 92.3%. The educational status of the respondents is that 61.9% of borrowers and 61.8% of non–borrowers are up to 5 years schooling and in total 61.9%. The demographic profile of the household members of the respondents showed out of 3610 in total showed that 35.5% falls between 0-15 following 28.80% between 16-30, 26% between 31-50 and 9.8% above 51 and their marital status above age 16 showed that out of 2329 members 65.7% is married, 28.6% un-married and 5.7% widows and separated. The educational statues of them above age 7 is that 52.1% has 1-5 years schooling following 19.9% 6-9 years, 15.7 with no education, 10.8% SCC and HSC and only 1.5% graduate and above. The occupational status of them between 16 and 60 aged is 32.3% household following day labor 13.8%, student 12.6%, others 12.6%, service/business 10.8% off-farm activities 10.4% and farming 7.4%. The income earner member above the age of 16 of the households showed that 57.3% members has no work while 30.30% full-time workers, 12.3% part-time workers and others 0.2%. The analysis of housing information of the households of the respondents is explored that 97.1% of borrowers, 93% of non-borrowers and in total 95.8% of both owns housing ownership. The data on possession of durable assets by households showed that among the top 5 assets 97.1 bed, following 96.8 households owns *Chauki* following, 91.7% mobile phone, 85.7% electric fan, 83.9% table/chair but only 19.5% holds television and 4.1% holds radio. (v) The land holdings by the households showed that 5.6% holds no homestead land, 79.5% holds 1 to 15 decimals, 7.4% holds 16-50 decimal and 7.4% holds more than 50 decimals. The data on cultivable land (own) explored that 71.1% landless following 15.4% holds more than 50 decimals, 6.7 holds 1 to 15 decimals also 6.7% holds 16-50 decimal. The holding status of cultivable Land (Leased-in or sharecropped) portrayed that 65.5% holds no land following 22.7% more than 50 decimal,7.4% 16-50 decimal and 4.3% 1 to 15 decimals. The data on holdings of pond land showed that 96.4% has no pond land while 3.6% holds 1 to 15 decimals. The data on non-land assets holds showed that 197 (41%) borrowers and 72 (32%) non borrowers hold livestock following cultivation instruments by 161 (33%) borrowers and 59 (26%) non borrowers, fishing net 68 (14%) borrowers and 43 (28%) non borrowers. The analysis of households' knowledge about micro-credit benefits showed that 95% respondents have knowledge about micro-credit benefit while 85.6% tried to get those benefits. For getting credit firstly 60.2% attempted to nongovernment (MFI/NGO/Insurance) following 13.6% local Money Lender (Mohajan/Private Samittee and 6.9% to more than one sources and 1.5% to government (Banks/Co-operatives). After first attempt the households attempted in second time for getting credit and 9% tried to local Money Lender (Mohajan/Private Samittee following 3.9% Nongovernment (MFI/NGO/Insurance), 2.8% Government (Banks/Co-operatives), 0.3% non-interest loan (Relatives/friends/neighbors) and only 0.1% to more than one sources. The respondents seek helped to get credit to different persons or bodies and among these 34.4% went to UP Office following 6.6% relatives/neighbors/friends, 5.5% NGOs, 3.4% government officer, 0.8% UP chairman and 0.7% to UP Member. Among the respondents 2.9% was asked to give money (bribe) for giving micro-credit benefit and 55% told that micro credit remove poverty. (vii) The top four causes out of fifteen for not getting credit: (i) no micro-credit in the area; (ii) non availability of collateral; (iii) misappropriation of credit and (iv) loss in business. The data on month of inclusion of credit showed that both formal and informal loans are taken during October to December (133+59) following January to March (86+108), July to September (64+32) and April to June (75+14). The rate of interest varied between 1% to 25% and more for both of formal and informal credits. The loans are paid in installment basis and the data showed that most of the loans of both formal (287) and informal (9) are on weekly basis following monthly formal (63) and informal (69) and biweekly, quarterly and annually are very small numbers while the number of installments is varied between 1 to more than 24 and the most of the loans of both formal and informal are in more than 24 (formal 290, and informal 12, and total 302) installments. The duration of loans is almost one year case of both formal and informal credits and paid against collateral. The payment of both types of loan is totally successful and both types are loans are paid regularly in time. All the loans are still going and remain a potion is unpaid. The six causes are identified as significant by PCA out of seventeen causes. The first dimension explained 15.14% variations and composed of purchasing of food items, repairing cost of houses, healthcare expenditure and education. The second dimension explained 9.14% variations with the factors of rearing cattle/poultry, purchasing of livelihood and payment of loan equipment. The third dimension is consisted of crop production and tackling shocks of natural calamities with explaining 8.3% variations. The fourth dimension has explained 7.29% variations composing the tackling shocks of sudden death of HH head and others. The fifth dimension is composed of Sending family member to abroad fish farming/fishing by explaining 7.93% variations. The sixth dimension has explained 6.43% variations having trade/business/industry and daughter/son's marriage purposes. The respondents were given to mark 24 heads of expenditure and investment out which fourteen heads were recognized and the top seven (7) heads in total of both formal and informal credit are consumption on food (151) following purchasing agricultural inputs (102) payment of loan (93), family enterprises (89), purchasing animals (83) housing improvements (72) and health care (71). (xi) The impact of informal micro-credit through economic performance based on before-after comparison showed that annual average total income has been increased from Tk. 87393.7 before receiving loan to Tk. 103145.7 after receiving loan than that of before significantly and with this line labor sale has been increased significantly from Tk. 54737.4 to Tk. 68075.8 and all the components of total income showed an increase after receiving the loan except business. The annual average total expenditure has been increased from Tk. 90948.8 before receiving loan to Tk. 116220.5 after receiving loan than significantly. Similarly, the information on impact of formal micro-credit through economic performance explored that: agricultural, non–agricultural and labor sale have been increased significantly though income from business and donation have been increased insignificantly. The amount of debt also increased without significance. The total expenditure and its two main components consumption and investment both have been increased significantly. The elements of investments: education & training, medical, agricultural, family business, household development and others have been increased significantly but savings increased insignificantly.

The impact of micro-credit in terms of formal and informal credit receiving households based on DID method showed the overall impact of both formal and informal credit. It portrayed that total income; total expenditure and investment have been increased 13.57%, 10.39% and 26.17%. All the elements of total income have been increased except debt which has been decreased by 2.39%. But the decrease in debt is the good sign of positive impact of debt. Consumption of food has been increased but non-food has been decreased. All the elements of investment have been increased except some factors. The savings has been decreased due excess increase in investment.

The percentage change of households' benefits showed that total income has been increased by 3.25% over the study period. All the elements of total income have been increased except labor sale while it has been decreased. It means that some of the borrowers have become self-reliance by using debt as debt has been also increased. The total expenditure has been increased by 8.76% over the study period. The consumption expenditure is one of the elements of total expenditure has been decreased by 4.42% while investment (capital) expenditure has been increased at a high rate (31.3%) as a result saving has been decreased by 11.8%. The percentage distribution of households' food insecurity during the study period showed that normal food insecurity of both borrowers and non- borrowers have been decreased during the study period and in total it has been decreased from 22.1% to 13% similarly no food insecurity of both borrowers and non-borrowers have been decreased during the study period and in total it has been decreased from 28.1% to 24.7% but finally no food insecurity has been increased of both borrowers and non-borrowers over the study period and in total it has been increased from 49.8% to 62.3%. Therefore, the use of micro credit ensures food security and reduces food insecurity. The percentage distribution of households self-assessed socio-economic status over the study period showed that the role of micro-credit graduation of households from lower class to upper class and the analysis of data  found that both borrowers and non-borrowers under extremely poor has been decreased over the study period and  in total it has been decreased from 25.5% to 13.9% similarly both borrowers and non-borrowers under moderately poor has been decreased over the study period and in total it has been decreased from 31.7% to 26.6% while both borrowers and non-borrowers under poor has been increased over the study period and in total it has been increased from 36.6% to 50.4% and both borrowers and non-borrowers under middle class has been increased over the study period and in total it has been increased from 4.9% to 7.9% and yet there is no change in rich category. The impact of micro-credit on changing educational and healthcare expenditure in 2019/20 compared to 2016/17 showed that education expenditure has been increased in case of both borrowers and non–borrowers over the study period and in total it has been increased from 18.4% to 58.8%. The data showed that health expenditure has been increased in case of both borrowers and non–borrowers over the study period and in total it has been increased from 29% to 57.6%. The analysis of factors responsible for non-payment of loan showed that that among the 15 factors rate of interest is very high is agreed by 64.2% of formal borrowers following 60.1% of installment period is very short, 50.8% of investment loss 47.8% of acute food problem and natural calamities 46.9% regarding informal credit while in case of formal credit 84.3% of rate of interest is very high following 77.2 of misappropriation of loan, 75.6% of installment period is very short, 72.4% of both acute food problem and investment loss and 63.8% of natural calamities. In total, rate of interest is very high stood (69.5%), installment period is very short is in second (64.1%) investment loss is in third (56.5%), acute food problem is in forth (54.2%), and medical treatment/medicine is in the fifth (49.5). Therefore, in case of both formal and in-formal cases the top most five factors of non-payment of loan are about same. The analysis of attitude towards micro-credit showed that among the 16 factors in case of formal credit among the top five factors showed that the rate of interest of micro-credit is reasonable is disagreed by 74.6% and agreed only by 17% following duration of credit is sufficient disagreed by 61.7% and agreed only by 21.2%, amount of credit is sufficient disagreed by 47.5% and agreed only by 35.5%, by micro-finance your savings has increased is disagreed by 37.4% and agreed only by 32.4% and terms and conditions are not rigid disagreed by 30.4% and agreed only by 57.5%. Side by side in case of informal credit portrayed that among the 16 factors the top five factors showed that the rate of interest of micro-credit is reasonable is disagreed by 94.5% and agreed only by 3.1% following duration of credit is sufficient disagreed by 78% and agreed only by 18.9%, amount of credit is sufficient disagreed by 77.2% and agreed only by 19.7%, by micro-finance your savings has increased is disagreed by 66.9% and agreed only by 9.4 % and terms and conditions are not rigid disagreed by 57.5% and agreed only by 37.5%. Over-all we found that the top five factors are the rate of interest of micro-credit is reasonable is disagreed by 79.8% and agreed only by 13.4% following duration of credit is sufficient disagreed by 66.0% and agreed only by 20.6%, amount of credit is sufficient disagreed by 55.3% and agreed only by 31.3%, by micro-finance your savings has increased is disagreed by 45.2% and agreed only by 26.4 % and terms and conditions are not rigid disagreed by 37.5% and agreed only by 30.9%. The analysis of causes of not over-coming from vicious cycle of poverty of not overcoming from vicious cycle of poverty showed that thirteen causes are listed and the respondents marked on three point scales which are agree, neutral and disagree. It is found that the most five causes agreed by non-borrowers are natural calamity (89%), following not getting loan in time (77.2%) loss of investment (75.9%), high interest (75.4%) and number of dependent members are high (69.7%). In the same way the most five causes agreed by borrowers are high interest (89.1%) following pressure of loan payment (76.5%), duration of loan is insufficient (71.5%), Insufficient loan (59%), natural calamity (58.8%) and loss of investment (58.1%).  In total the top five agreed causes of not over-coming from vicious cycle of poverty are high interest (84.7%) following pressure of loan payment (71.2%), natural calamity (68.4%), duration of loan is insufficient (64.7%) and loss of investment (63.8%). The findings of FGD explored that taking informal credits the total income, total expenditure comprising of consumption and investment both has been increased simultaneously with income. The investment in agricultural and productive has been increased highly. The savings has been decreased as the investment increased with income. The socio-economic impact of formal micro-credit: revealed that the agricultural and non-agricultural income and labor sale have been increased with the increase of loans. The consumption of both food and noon food items has been increased while savings increased slightly. The causes of non-payment of loan in time are rate of interest, short of installment period, investment loss acute food problem and natural disasters are the main causes of non-payment of formal loans while in case of informal loans the causes are almost same of non-payment of loans. Finally, the study suggested that for breaking vicious cycle of poverty by micro-credit the duration of loans should be at least five year and the volume of loans must be minimum 500,000 and repayment should at not be less than monthly. The rate of interest should not be more than 5%.

# List of Acronyms

| | |
|---|---|
| BBS | Bangladesh Bureau of Statistics |
| BHWDB | Bangladesh *Haor* and Wetland Development Board |
| CBN | Cost of Basic Needs |
| CEGIS | Center for Environment and Geographic Information Services |
| DCI | Direct Calorie Intake |
| DID | Difference-in-difference |
| FEI | Food Energy Intake |
| FGD | Focus Group Discussions |
| GoB | Government of Bangladesh |
| HH | Household |
| HIES | Household Income and Expenditure Survey |
| LPL | Lower Poverty Line |
| LPL | Lower Poverty Line |
| MAS | Multi-agent systems |
| MC | Micro-credit |
| MDGs | Millennium Development Goals |
| MFI | Micro Finance Institution |
| MPI | Multidimensional Poverty Index |
| NGO | Non-Government Organization |
| PCA | Principal Component Analysis |
| SACCOS | Savings and Credit Co-operative Societies |
| SFYP | Seventh Five Year Plan |
| SMCP | Savings and Micro Credit Program |
| SPSS | Statistical Package for Social Sciences |
| UP | Union Parishad |

# CHAPTER ONE
# INTRODUCTORY ASPECTS OF THE STUDY

## 1.1 An Overview of *Haor* Area in Bangladesh

*Haors* are bowl-shaped large tectonic depressions which receive surface runoff water and consequently become a very expensive water bodies in the monsoon and mostly dries up during post-monsoon period. *Haor* districts (Sunamganj, Sylhet, Habiganj, Maulvibazar, Netrakona, Kishoreganj and Brahmanbaria) of Bangladesh cover 19,998 sq. km land, which is 13.56% of total area of the country. According to Center for Environment and Geographic Information Services (CEGIS) the total land of the *haor* districts is about 43% (8585 sq. km) is under wetland (*haor*) where 373 *haors* are exists (CEGIS, 2012). The statistics of *Haors* in Bangladesh is given in Table-1.1

**Table 1.1: Distribution of *Haor* Areas in Bangladesh.**

| Districts | Total area in ha (% of total) | *Haor* area in ha (% of total) | No. of *haor* (% of total) |
|---|---|---|---|
| Sunamganj | 367,000 (18) | 268,531 (31) | 95 (25) |
| Sylhet | 349,000 (17) | 189,909 (22) | 105 (28) |
| Habiganj | 263,700 (13) | 109,514 (13) | 14 (4) |
| Maulvibazar | 279,900 (14) | 47,602 (6) | 3 (1) |
| Netrakona | 274,400 (14) | 79,345 (9) | 52 (14) |
| Kishoreganj | 273,100 (14) | 133,943 (16) | 97 (26) |
| Brahmanbaria | 192,700 (10) | 29,616 (3) | 7 (2) |
| Total | 1,999,800 (100) | 858,460 (100) | 373 |

*Source:* Master Plan of *Haor* Area, Volume,1, Summary Report, Government of the People's Republic of Bangladesh Ministry of Water Resources Bangladesh *Haor* and Wetland Development Board. p.1.

The *haor* region has long been lagging behind the main stream of the national development due to geographical location and considering this, in 1974, then the Government of Bangladesh (GoB) established an independent Board for the development of the *Haor* region (CEGIS, 2012). Following this establishment of the Board, the GoB established the Bangladesh *Haor* and Wetland Development Board (BHWDB) through a resolution approved by the Cabinet in 2000. The mandate of the BHWDB is to coordinate the activities and formulate projects relating to a holistic development of the *Haor* and wetlands of the country (CEGIS, 2012). The map of *haor* regions is given in Figure-1.1.

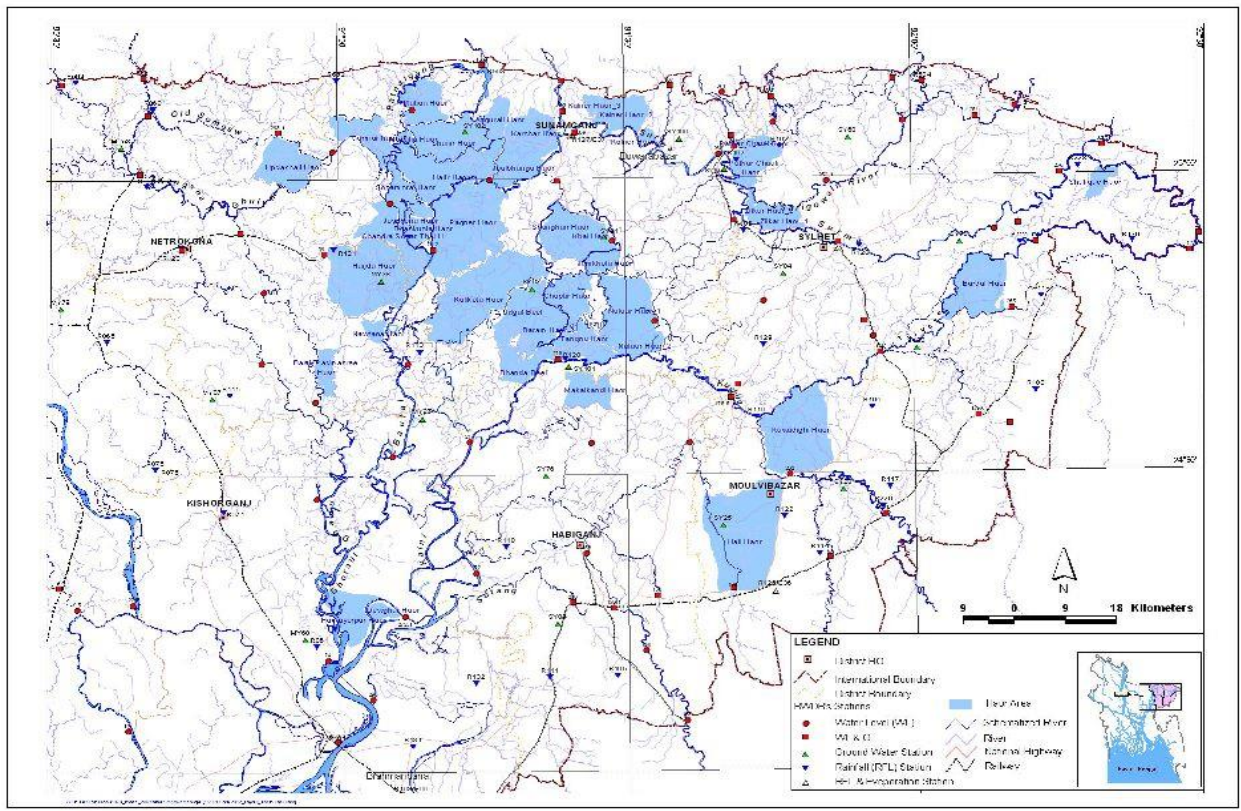

**Figure 1.1:** *Haor* **Areas of Bangladesh.**

## 1.2 Sunamganj District

Sunamganj district is situated in the North-East region of Bangladesh. Geographically its location is 25.03869 degree North and 1.403761 degree East. The total area of the district is 367,000 hectors of which 268,531 hectors are wetland *i.e.*, *haor* which is 31% (Table-1). Therefore, about one-third of total area is *haor* and that is why it is called the most *haor* based district in Bangladesh where 95 *haors* are located (Table-1).  In terms of kilometer total area of the district is     3,669.58 sq.km. The total population of the district is 2,467,968 according to census 2011 but according to 1.25% growth rate on average it would be now around 27, 00,000. In Sunamganj district there are 11 (eleven) upazilas, 4 municipalities, 36 wards, 139 mahallas, 82 union parishads, 1711 mouzas and 2813 villages. The upzilas are Bishwamvarpur, Chhatak, Derai, Dharmapasha, Dowarabazar, Jagannathpur, Jamalganj, Daskin Sunamganj, Sullah, Sunamganj Sadar and Tahirpur.  The map of Synamganj District is given in Figure-1.2.

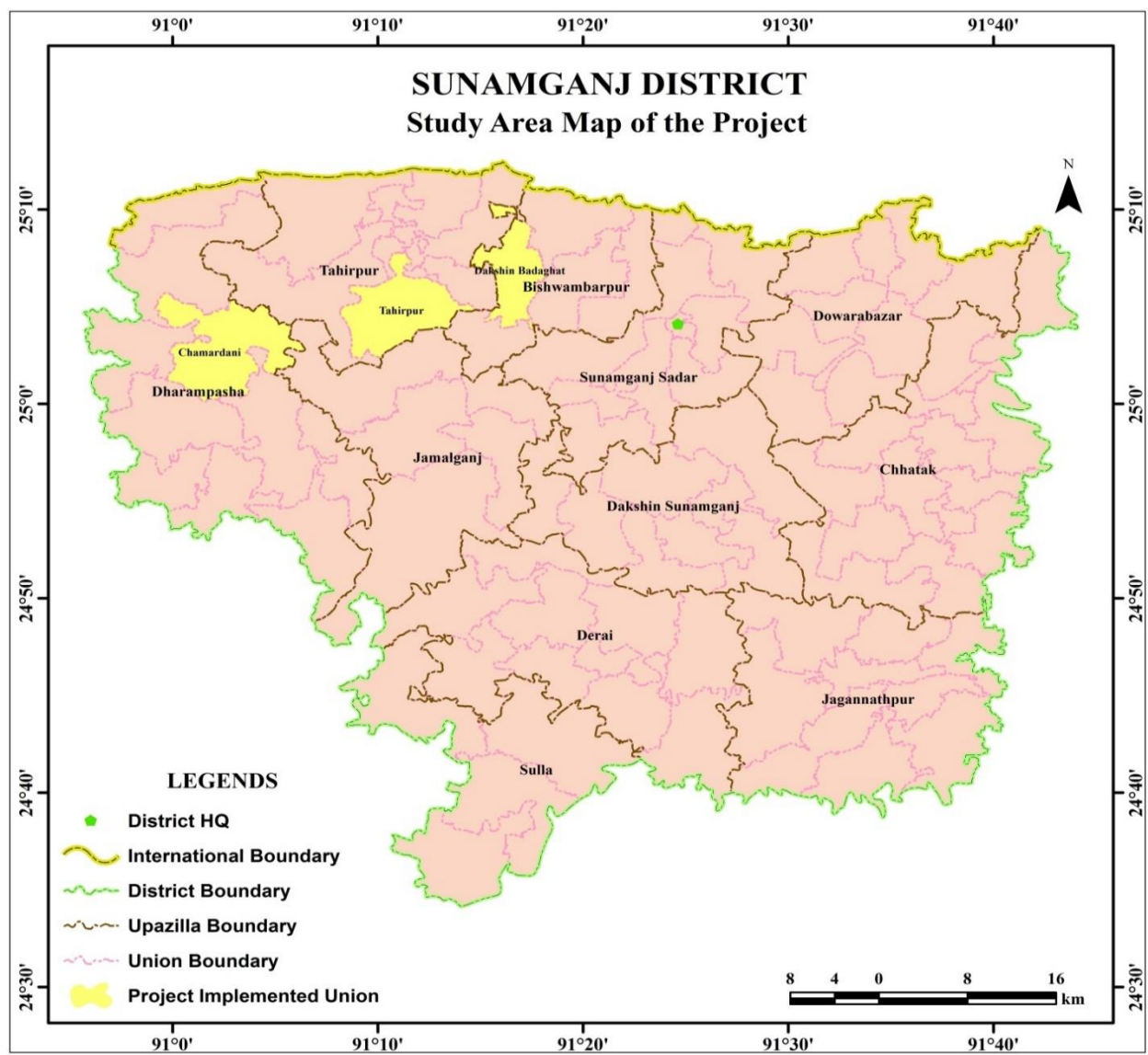

**Figure 1.2: Sunamganj District.**

## 1.3 Poverty in Bangladesh

Poverty alleviation is considered to be one of the most important indicators of the socioeconomic development of a state and society. Bangladesh has achieved remarkable development in poverty alleviation during the last few decades as a result of the combined efforts of both the Government and non-government sectors. According to the 'Millennium Development Goals: End period Stocktaking and Final Evaluations Report' the incidence of poverty has declined 1.74 percentage points on an average in Bangladesh during 2000-2010 against the MDGs target of 1.20 percentage points. According to the recently published 'Household Income and Expenditure Survey 2016' the present poverty rate is 24.3 percent whereas it was 56.7 percent in 1991. The Government has set up a target to reduce the poverty to 18.6 percent at the end of the 7th Five Year Plan (2016-2020). Despite all these positive changes in poverty reduction, still one-fourth population of Bangladesh lives below the poverty line. It would not be possible to attain the desired level of socioeconomic development without emancipating this portion of population from poverty. For this reason, the Government still considers poverty alleviation as a major agenda on the policy and development issues of the country.

Bangladesh has achieved a significant progress in the Human Development Indicators. According to the UNDP Development Report-2016 the position of Bangladesh has been ranked at 139-th among 187 countries. Furthermore, the report reveals that Bangladesh's Multidimensional Poverty Index (MPI) reduced to 0.188 in 2016 from 0.237 in 2007.

**1.3.1 Measurement of the Incidence of Poverty in Bangladesh**

The measurement of the incidence of poverty in Bangladesh is taken from the report of Household Income and Expenditure Survey (HIES). At present, the survey is renamed as Household Income and Expenditure Survey (BBS,2016). The first HIES in Bangladesh was carried out in FY 1973-74 and  after that up to FY 1991-92, few HIESs were conducted maintaining the same strategies of the first one. HESs were accomplished by Food Energy Intake (FEI) and Direct Calorie Intake (DCI) method. According to this survey, a man having calorie intake of less than 2,122 kilo-calories daily to be considered as absolute poor. On the other hand, a man having an intake of below 1,805 kilo-calories is considered as hard-core poor. The Bangladesh Bureau of Statistics (BBS) has adopted 'Cost of Basic Needs (CBN)' for HIES for the first time in 1995-96. The same method applied in the HIES in 2000, 2005 and 2010. This method also considered non-food consumption items for compiling poverty index. The latest HIES has been undertaken in 2016 and recently its result has been published.

**1.3.2 Trends of Poverty**

The incidence of income poverty (measured by CBN considering the upper poverty line) declined nearly 7 percent (from 31.5 percent to 24.3 percent) over the period in 2010-2016. During this period, the compound poverty declined 4.23 percent annually. On the other hand, the rate of income poverty declined from 40.0 percent to 31.5 percent from 2005 to 2010. At that time, compound poverty got reduced by 4.67 percent each year. Therefore, it is evident that though poverty is decreasing gradually, the pace of reduction rate declined during the period of 2010-2016 compared to the period of 20052010. In urban areas poverty reduction rate is higher (4.68 %) than rural areas (1.97%). During 2010 to 2016, the reduction rate of the depth of poverty (measured by poverty gap) was 4.28 percent. It has also been observed that income poverty reduction rate in urban areas is lower than that of rural areas (1.61% and 5.12% respectively). Moreover, the reduction rate of the depth of severity of poverty (measured by squared poverty gap) was also lower in urban areas compared to rural areas. The trends of poverty are depicted in Table 1.2. The HIES-2016 reveals that on the basis of lower poverty line poverty rate in 31 districts is above national average. On the other hand, using the upper poverty line poverty rate in 36 districts is above national average.

**Table 1.2: Incidence of Poverty in Bangladesh.**

|  | 2016 | 2010 | Annual Change (%) (2010 to 2016) | 2005 | Annual Change (%) (2005 to 2010) |
|---|---|---|---|---|---|
| **Head Count Index** |  |  |  |  |  |
| National | 24.3 | 31.5 | -4.23 | 40.0 | -4.67 |
| Urban | 18.9 | 21.3 | -4.68 | 28.4 | -5.59 |
| Rural | 26.4 | 35.2 | -1.97 | 43.8 | -4.28 |
| **Poverty Gap** |  |  |  |  |  |
| National | 5.0 | 6.5 | -4.28 | 9.0 | -6.3 |
| Urban | 3.9 | 4.3 | -1.61 | 6.5 | -7.93 |
| Rural | 5.4 | 7.4 | -5.12 | 9.8 | -5.46 |
| **Squared Poverty Gap** |  |  |  |  |  |
| National | 1.5 | 2.0 | -4.68 | 2.9 | -8.81 |
| Urban | 1.2 | 1.3 | -1.33 | 2.1 | -8.64 |
| Rural | 1.7 | 2.2 | -4.21 | 3.1 | -8.75 |

**Source:** BBS, HIES-2016

## 1.4 Statement of the Problem

There are a number of studies observing the positive impact of micro-credit (MC) on clients with regard to material well-being, reduction in exposure to seasonal and climate change vulnerability (Finance Division, 2014), contributions to consumption smoothing, and a better ability to deal with crises (Khanom, 2014; Mamun et al., 2013; Khalily, 2004; Khandker 1998, Zaman, 2001, 1998; Mustafa et al. 1996). However, few of these studies compare results with those not members of MC programs. But there are some studies evidencing that microfinance increase vulnerability of the poor and they argued that the measuring the wellbeing from micro credit impact on individual borrower should be determined (Islam, 2014). The diversification of micro credit sometimes leads to the decline of economic development and poverty reduction (Elahi & Danapoulos, 2004). There are numerous problems in MC viz. the different interest rate is charged by different MCIs. The present study will analyze the impact of MC program considering both the clients and non-clients of MC which had been followed by latest researchers (Habib & Jubb, 2015; Islam, 2014; Khan, 2013).

Sunamganj district is one of the *haor* based and poorest district in Bangladesh (SDC, 2006). People living in this area are considered as "backward section of citizens. A number MC Institutions have distributed a large volume of microcredit to improve the socio-economic conditions of this area and also, they are still doing for example BRAC solely disbursed loans among 51760 borrowers through its 39 branches up to December 2015 (BRAC, 2016). The research findings revealed that MC played a significant role in socioeconomic development of *haor* areas, especially household income increment, livelihood diversification, creation of self-employment, poverty reduction and women empowerment though it entrapped few households in vicious cycle of poverty (BRAC, 2015; IFAD, 2015). A number of development projects have also been implemented to improve the socio-economic condition of this area viz. SCBRMP by government and many international donors and agencies viz. IFAD, SDC, CARE etc. But the area is still lagged behind from others parts of the country in terms of education, poverty and others benefits as stated in Table 1.3:

**Table 1.3: Cooperative Development Indicators of Bangladesh and Sunamganj District.**

| Factor | Bangladesh | Sunamganj |
|---|---|---|
| Education | 70% | 35% (lowest) |
| Poverty rate | 25% | 36% |
| Monthly per capital Income | | 3rd lowest |
| Monthly per capita consumption | | 11th lowest |
| Under-five mortality rate 2009 (per 1000 live birth) | | 94 (3rd highest) |
| Education development index (EDI) | | 3rd lowest |
| Roads and Electricity | | 2nd lowest |

*Source:* Lagging Districts Development, Background Study Paper for Preparation of the
Seventh Five-Year Plan by Bazlul Haque Khondker, Moogdho Mim Mahzab

Therefore, there is a huge disparity in development of Sunamganj district in terms of national index. The causes behind this lagging may be numerous but one of the causes may be the ineffectiveness of micro-credit (MC) program as MC improves socioeconomic condition of nationally and internationally as stated above. The studies on the impact of MC on *haor* area mostly considered the poverty alone but yet there is no study covering the two or more socioeconomic indicators of clients and comparing clients versus non-client. The present study will measure the impact of MC programs comparing different socioeconomic aspects of both the client and non-client concurrently in order to measure productive outcome of MC programs.

## 1.5 Objectives of the Study

The core objective of project is to assess the impact of micro credit on the livelihoods of the clients in the *haor* area of Sunamganj district. The specific objectives are to:

i.   identify the nature of microcredit services available in the *haor* area of Sunamganj district;

ii.  measure the productive outcome of microcredit on the livelihoods of the clients;

iii. identify the problems in microcredit facing by the clients in the utilization of microcredit services; and

iv.  draw implications for design and implementation of micro-credit programs in the *haor* areas of Bangladesh specially for Sunamganj district

## 1.6 Research Questions and Corresponding Expected Results

Before identify the expected results from the study we have to recognize the questions to be answered through this study. So, the research questions and their corresponding results are given in the following Table 1.4 in line with the research objectives:

**Table 1.4: Research Objectives, Research Questions and Corresponding Expected Results.**

| No. of Specific objectives | Research questions behind expected results | Corresponding Expected results of research questions |
|---|---|---|
| (i) | ❖ What are the different MC programs in the *haor* areas? <br> ❖ What are the different MC programs run by the government in the *haor* areas? <br> ❖ What are the different MC programs run by the non-government/donors /agencies in the *haor* areas? | Understanding the nature and features microcredit programs in the *haor* area especially in Sunamganj district. |
| **(ii)** | ❖ Do MC programs have a positive impact on livelihoods of the clients in terms of selected social indicators viz. income, consumption, assets, net worth, education, access to finance, social capacity, food security and handling socks etc.<br>❖ Whether the MF clients are improving more their socioeconomic condition than non-clients. | Impacts of micro credit on clients' livelihoods;<br><br>Impacts matrix |
| **(iii)** | ❖ Whether the cost of microcredit is higher than the rate of internal return of it or not?<br>❖ Whether the volume of microcredit is insufficient in terms of productive investment or not?<br>❖ Whether the duration of credit is shorter than the client's need for productive outcome or not?<br>❖ Whether the payment schedule of credit is tighter than the expectations of clients?<br>❖ Whether the continuation of credit is enough or not for productive outcome?<br>❖ Whether the availability of credit is limited or not?<br>❖ Whether the volume of credit is limited or not? | Challenges and problems that are facing by the MC clients.<br><br>The problems matrix of MC. |
| **(iv)** | ❖ What are the policy recommendations and actions should take to enhance the productive outcomes from the MC programs?<br>❖ How to implement these recommendations, adjustment and coordination mechanism of MC programs in the *haor* areas? | ✓ Recommendations, policy options, adjustments and action for future plan for the productive outcome of MC programs in the *haor* areas. |

## 1.7 Relevance of the Project

As stated earlier, the microcredit programs played important roles in increasing income, assets, net worth, education, financial capacity, consumption etc. The *haor* based Sunamganj district is far behind from the national level in terms of different socioeconomic indicators. Many microcredit institutions are working to improve the socioeconomic conditions of this area for a long time. But still the socioeconomic situations of people of this area are not up to the national level. The findings of the study would help to improve the effectiveness of microfinance programs of this area which would reduce regional disparities and help to achieve the targets of the national Seventh Five Year Plan (SFYP) as stated in Table 1.5:

**Table 1.5: Targets in Seventh Five Year Plan to Achieved through the Study.**

| Targets:<br>Production, Income Generation<br>and Poverty | Current<br>Situation<br>(2005-2010) | Vision<br>2021 | Progress<br>Six FYP<br>2015 | Target in<br>Seventh<br>Five-<br>year plan |
|---|---|---|---|---|
| 1. Real GDP Growth (%) | 6.1 | 10 | 6.5 | 8 |
| 2. Head Count Poverty (%) | 31.5 | 13.5 | 24.8 | 18.6 |
| 3. Reduction of Extreme Poverty (%) | 17.6 | 0 | 12.9 | 8.9 |
| 4. Share of Manufacturing Employment (%) | 12.4 | 20 | 15.4 | 20 |
| 5. Net Enrolment at Primary Level (% | 843 | 2000 | 1314 | 2009 |
| 6. Net Enrolment at Primary Level (%) | 91 | 100 | 97.3 | 100 |
| 7. Net Enrolment at Secondary Level (%) | 43 | 100 | 57 | 100 |
| 8. Net Enrolment at Tertiary Level (%) | 9 | 20 | 12 | 20 |
| Cohort | 35 | 20 | 50 | |

Therefore, the study is completely with the line of SFYP that must help to achieve the targets of SFYP and ultimately the Sustainable Development Goals will be achieved. Further, the study will help to formulate policy by the concerned authorities to reduce poverty and sustainable development in the *haor* areas of Bangladesh and especially for Sunamganj district. The authorities of micro finance bodies will make or revise their terms and conditions of micro credit, volume and payment structure friendly for clients to increase productive outcome.

# CHAPTER TWO
# REVIEW OF RELATED LITERATURE

Review is essential because of to establish the linkage of the present study with the previous studies, identify the uniqueness of the present study, find out the appropriateness of the methodology used in the study and lastly to justify the plan and preparation to complete study we proposed.  With this mind, the   researcher has reviewed the studies made by different scholars in this field from the following aspect viz. (i) studies made internationally on microcredit and poverty alleviation; (ii) studies regarding Bangladesh on microcredit and poverty alleviation and (iii) studies regarding Sunamganj district on microcredit and poverty alleviation. These are presented below:

## 2. Literature Review

There are a number of studies (ALAM & HOSSAIN, 2018; Schreiner Mark, 2001; Mazumder Mohummed Shofi Ullah and Wencong Lu, 2013; Bangladesh Bank, 2015; Rahman Sayma, 2007; Habib and Jubb, 2015; Khandker Shahidur, 2003, 2005; CGAP, 2010; Chowdhury M. Jahangir Alam et al., 2002; Rabby Talukder Golam, 2012; Khanom Nilufa A. ,2014; Zama Hassan, 2001; Ahmed Salehuddin, 2017; Uddin Mohammed Salim, 2011; Shoji Masahiro, 2008; Fant Elijah Kombian, 2010)   recognized that  benefits of  micro credit exceeds cost and it  instilled borrowers income, assets endowment, standard  of living and poverty reduction, productivity of business  and agriculture, increase wealth ,savings ,local economy , consumption stable finance major expenses and cope with shocks, food intake  and increase in  sustainable community acquisition of asset. Anthony Denise (2005) found that microcredit as competing mechanisms which have differential effects group identity, sanctions and reciprocity are all associated with more borrowing in the group.  Micro-credits make smooth consumption and enables savings and income decisions (Lawrence Peter, 2004).  Alam Saad (n.d) showed that micro-credits enhance self-employment profit by 50% to 81%.  Micro-credits boost up credit facilities (Ogaboh et al., 2014) and   heighten repayment performance of borrowers Barboza and Barreto,2006, Giné and Karlan (2007, 2009) also amplify 'self-help' approach for development (Rankin Katharine N., 2001). The long- run effects of micro-credit reduce inequality in the society (Ahlin and  Jiang, 2005)  and make free poor people from the curse of poverty (Khandakar Shahidur R., 1998)   Khan Mohammad Mohaiminuzzaman (2013) in his master found that green microcredit is becoming more popular among the natural resource-dependent borrowers in Bangladesh and the borrowers get training facilities   Karlan andValdivia, 2006) in household level micro credits play positive role in   risk management (Karlan  and Zinman, 2009). Furthermore, Alam & Zakaria (2021) showed a positive relationship between a threshold income level and consumer's awareness towards green environment, where an increase of their marginal income plays a key role through an income generating process towards the urban bases of environmental awareness. Midgle James (n.d.) found that microcredit and microfinance programs have helped to create small businesses, or otherwise strengthened existing small businesses in low-income communities.  Bylander Maryann (2014) found that using micro credit in combination with migration allows households to immediately meet consumption goals and utilize the credit being actively promoted by microfinance institutions, while also retreating from insecure and less profitable local livelihood strategies. Diagne Aliou (1999) found that a higher share of land and livestock in the total value of household assets is negatively correlated with access to formal credit. However, land remains a significant determinant of access to informal credit. Angelucci et al. (2013) found that the average effects on a rich set of outcomes and suggest no transformative impacts, but more positive than negative impacts. Islam Khan Jahirul (2014) argued that to get positive impact from micro-credit in reduction of poverty the nature and term should be dynamic with the nature of poverty. Mwangi and Sichei (2018) made a comparative study between the period 2006 and 2009 on access to credit of Kenyans using multinomial probit models and found that increase in household size reduced access to bank loans and ASCAs while it promoted access to loans from buyers of harvest. Increase in distance to service provider led to a decline in access to credit even though the impact was marginal. On the other hand, increase in age; education and income tend to enhance access to credit but the probability of access drops as one draws close to retirement age. Banerjee et. al (2011) reported the results of a randomized impact evaluation of a program designed to reach the poorest of the poor and elevate them out of extreme poverty and revealed that the program results in a 15% increase in household consumption and has positive impacts on other measures of household wealth and welfare, such as assets and emotional well-being. Morduch Jonathan (1999) experienced that many relatively poor households can save in quantity when given attractive saving vehicle.

 Micro-credits play positive role in empowering women (Rahman Sayma,2007; Khandker Shahidur, 2005; Omorodion, 2007). Pitt Mark M. and Khandker Shahidur R. (1998) found micro-credit program has a larger effect on the behavior of poor households in Bangladesh when women are the program participants then men. For example, annual household consumption expenditure had increased by Tk. 18 for every Tk. 100 additional borrowed by women from these credit programs compared with 11 takas for men. Ahmed Ferdoushi (2011) also showed that, the 'with credit' women have a much lower percentage of poverty in terms of its incidence (80%), intensity (28%) and severity (12%) compared to the 'without credit' respondents (99, 59 and 37% respectively) and also found that educational attainment of the respondents and income earners in the family contribute positively to reduce poverty situation among the 'with credit' households more, as compared to 'without credit' households and thus he concluded that microcredit program helps the rural women to reduce their poverty more effectively. Abdalla Nagwa Babiker (2009) found that majority of women in Sudan live with low or no income; economically they are dependent on their husbands' income; burdened with their household activities and responsibilities to feed; educate and take care of many children, encounter a core problem which is lack of access to credit and financial services to economically, socially and politically empower themselves and improve their status. Rahman Aminur (1998) indicated that worker and borrowing peer loan group members in centers press on clients for timely repayment, rather than working to raise collective consciousness and borrower empowerment as envisaged in the Bank's public transcript. Mwongera Rose K. (2014) identified that majority of the young women entrepreneurs had borrowed money from the nearby micro-finance institutions the loan accrues interest rate imposed by the financial institutions as well as demand for collateral security and they found it unreasonable he also found out that most of the entrepreneurs had attained secondary school certificate as their highest academic qualification and was a determinant for uptake of loans and established that licensing of more financial institutions would encourage uptake of loan among women entrepreneurs since uptake of loan is low and this would influence business performance. Noreen Sara (2011) has made an attempt to explore the socio-economic determinants of women empowerment in which microfinance is crucial economic determinant. The results showed that women empowerment is considerably influenced by age, education of husband, father inherited assets, marital status, number of sons alive and microfinance. Age, education of husband, no of live sons and father inherited assets are more statistically significant variables in this study. Bhattacharjee Priyanka (2016) unveiled that most women failed to understand the process and effectiveness of microcredit programs and took loan for meeting the cash-demand of the male(s) within family as the feel microcredit as a medium to fulfill their emergency requirements. She also found that lack of education, awareness, unwillingness to join other programs of microfinance institutions, pessimistic thinking about microcredit programs, hostile family structure, negligent attitude towards repayment of loan(s), limited investing opportunities, etc. are the main causes that hinders the road to development. Afrin Sharmina et al (2008) showed that the financial management skills and the group identity of the women borrowers have significant relationship with the development of rural women entrepreneurship in Bangladesh and the experience from the parent's family of the borrowers and the option limit may also lead to the rural women borrowers to be entrepreneurial. Mohamed Fauzia Mtei (2008 argued that women and micro credit agencies have divergent understandings of money and its investment and its role in poverty reduction. Simojoki Hanna-Kaisa (2003) examined the socio-economic impact of micro-finance on urban female micro-entrepreneurs in Nairobi, Kenya and found that micro-credit plays crucial role in empowering women in business control and decision making. Adeyemo Comfort Wuraola (2014) examined the influence of vocational skills acquisition and micro-credit loans on widows socio-economic and psychological adjustments in South-western Nigeria and found that vocational skill acquisition centers afforded widows the opportunities to share their pains and experiences, thus assisting them to reduce loneliness, frustration, health-related problems and adjust to the reality of spouse's death. Vocational skill acquisition and micro-credit loans considerably assisted widows in overcoming their socio-economic hardship and psychological challenges. Karubi Nwanesi Peter (2006) established that micro-credit provides finance to enhance market and rural women's participation in production and trade. The study further ascertains that woman have some control over their loans. Churk Josephine Philip (2015) examined the Contribution of Savings and Credit Co-operative Societies (SACCOS) on promoting rural livelihood revealed that, SACCOS have played minimal role towards promoting rural livelihoods in the study area. Waliaula Rael Nasimiyu (2012) conducted the study to evaluate the relationship between microcredit and the growth of SMEs in Kenya and found that there was a very strong positive relationship between the variables. Gomez Rafael (2001) examined the determinants of self-employment success for microcredit borrowers and found that social capital is a positive determinant of self-employment earnings and neighborhoods play in fostering social capital and improving micro-entrepreneurial performance. Asgedom and Muturi (2014) analyzed the socio-economic factors that affect the institution's loan repayment performance and revealed that the level of education, loan amount and loan category have insignificant effect on the probability of the Savings and Micro Credit Program (SMCP) loan repayment. On the other hand, age, gender, type of business, origin of ethnicity (Minar and Abdul, 2020), and credit experience are significant determinants where age and type of business have negative relationship and gender and credit experience have positive relationship with the loan repayment probability.

On the other hand, there are many scholars (Rogaly Ben ,1996; I.G. Okafor,2016)  who have found that micro credit programs have negative and insignificant role on poverty reduction. Matsuyama Kiminori (2006) found that a movement in borrower net worth can shift the composition of the credit between projects with different productivity levels and showed how investment-specific technological change may occur endogenously through credit channels. Furthermore, such endogenous changes in investment technologies in turn affect borrower net worth. These interactions could lead to a variety of nonlinear phenomena, such as credit traps, credit collapse, leapfrogging, credit cycles, and growth miracles in the joint dynamics of the aggregate investment and borrower net worth. Osmani and Mahmud (2015) found that the impact of microcredit on the lives and livelihoods of the poor remains a contentious issue, there is no disputing the fact that there is something novel, something special, about microcredit that has allowed an altogether new mode of financial intermediation to emerge, providing credit to millions of hitherto 'un-bankable' poor without breaking the lender's back. That is an extraordinary achievement in itself. Nissank (2002) examined economics of microfinance as an instrument of microenterprise development and poverty reduction as well as its delivery mechanisms and also assessed empirical evidence of the performance of microfinance institutions and their impacts on poverty alleviation and microenterprise development. The donors target might have unintended negative effects on the very objectives of poverty reduction and microenterprise development. Ahlin Christian(2009) extended  the theory of micro credit movement  management  to examine sorting when group size is larger than two and joint liability can take several forms and  found that  moderately homogeneous sorting by risk, in support of Ghatak's theory and proved  the  evidences  of risk anti-diversification within groups while the anti-diversification results revealed   a potentially negative aspect of voluntary group formation and point to limitations of microcredit groups as risk-sharing mechanisms. Fasorant M.M. (2010) showed that the incidence of poverty was high among the economically active age bracket as the mean age was 33 years of the borrowers and   all respondents acquired formal education as 60% had above primary school education and 39.2% of total respondents had no specific occupation before the inception of micro-credit scheme.

Mamun et al. (2013) argued that microfinance has developed innovative management and business strategies by mobilizing the savings, loan repayment and empowerment of women, but its impact on poverty reduction remains in doubt. It certainly plays an important role in providing safety-net and consumption smoothening but identifying how microfinance can be used as an important vehicle to make an even larger and more critical contribution to alleviating poverty is in need of more careful assessment. Banerjee et. al (2015) showed the causal evidence on microcredit impacts informs theory, practice, and debates about its effectiveness as a development tool and found that a consistent pattern of modestly positive, but not transformative, effects. Angelucci et al. (2013) showed that the average effects on a rich set of outcomes measured of micro-credit suggest some good and little harm but estimators identify heterogeneous treatment effects and effects on outcome distributions, but again yield little support for the hypothesis that microcredit causes harm. Khandker Shahidur R. et al. (2014) examined the level of indebtedness of the micro credit borrowers in Bangladesh some borrowers might be taking loans that they will not be able to repay. He found that 26 percent of microcredit borrowers are over-indebted while 22 percent of non-microcredit borrowers. Cons and Paprocki (2010) opined that microcredit and other 'self-help' development strategies operate through idealized notions of poverty and rural life. The scholars explored the ongoing debate over microcredit in his study area and reflect on how re-rooting debates over development in specific places might move such debates from questions of 'self-help' to grounded and historicized projects of self-determination. Selinger Evan (2008) remarked on the debates about the Grameen Bank's micro-lending practices depict participating female borrowers as having fundamentally empowering or disempowering experiences and found that this discursive framework may be too reductive: it can conceal how technique and technology simultaneously facilitate relations of dependence and independence; and it can diminish our capacity to understand and assess innovative development initiatives. Babajide Abiola (2012) found that access to microfinance does not enhance growth of micro and small enterprises in Nigeria. However, other firm level characteristics such as business size and business location, are found to have positive effect on enterprise growth. Karlan and Zinman(2011) study found that microloans increase ability to cope with risk, strengthen community ties, and increase access to informal credit and therefore microcredit here may work, but through channels different from those often hypothesized by its proponents.

A few scholars have come to find the impact of micro credit on hoar area of Bangladesh. Among them Kazal et al. (2010) found the impact of micro credit on reduction of poverty and found the 2/5 of micro-credit receivers could not repay loan and fall in vicious cycle of credit. According to Khondker Bazlul Haque and Mahzab Moogdho Mim (laaging district dev.) the monthly per capita income is the third lowest of Sunamganj district which is only 2156 while the per capita monthly expenditure is 1978 the 11[th] lowest among the 15 lagging districts in Bangladesh. According to CGAP (2010) micro finance help hundreds of millions of people keep their consumption stable, finance major expenses, and cope with shocks despite incomes that are low, irregular, and unreliable in hoar area of Bangladesh. Planning Commission (2016) recommended from the workshop on "Environment and Climate Change Policy Gap Analysis in *Haor* Areas" that *haor* area is enriched with valuable aquatic flora and fauna including different species of fish, other natural resources and natural habitat. The *haor* basin has its ecological and economic importance and plays a vital role in the growth of the country but it is urgent to incorporate the economic importance on the Hoar areas in the national curriculum, particularly for the pupils of the hoar districts, so that the new generation can understand their economic and environmental significance.

There two types of micro credit viz formal and informal and there is difference of impact between these two types. Gichuki et al. (2014) found that the key challenges hindering micro and small enterprises from accessing credit facilities to be high cost of repayment, strict collateral requirements, unwillingness of people to act as guarantors, high credit facilities' processing fees and short repayment period in case of formal micro- credit hence they recommended that financial institutions set more flexible, affordable and attractive requirements in financing micro and small enterprise while Kazal et al., (2010) found that in the *haor* area informal credit is more stronger than formal sector. Barnaud et al. (2008) explored the Multi-agent systems (MAS) and the results of a series of simulations exploring the ecological, social and economic effects of various rules for formal and informal credit suggested by the villagers-participants. The study highlighted the ability of MAS to deal with interactions between social and ecological dynamics and to take into account social interactions, in particular the concept of social capital which is a determining factor when dealing with sustainability issues. Secondly, the study addressed the potential and limits of MAS models to support a bottom-up (or participatory) modeling approach. This experiment suggests that the usefulness of models relies much more on the modeling process than on the model itself, because a model is usually useless if it is misunderstood by its potential users, or if it does not respond to their current preoccupations. The intuitive representation of real systems provided by MAS and their high flexibility are the two underlined characteristics favoring their appropriation by local stakeholders. Phan Dinh Khoi (2012) in his PhD investigated the determinants of households" borrowing decisions and showed that informal microcredit alters the households" decisions to obtain a formal microcredit

# CHAPTER THREE
# METHODOLOGY OF THE STUDY

## 3.1 Study Area

A *haor* is a wetland ecosystem located in the north eastern part of Bangladesh which physically is a bowl or saucer shaped shallow depression, also known as a back swamp. There are about 373 *haors*/wetlands located in the districts of Sunamgang, Habigang, Netrakona, Kishoreganj, Sylhet, Maulvibazar and Brahmanbaria. There are 47 major *haors*, and 6 of them are ranked as internationally important wetlands. There are 6300 *Beel*, and 3,500 of them are permanent and 2,800 of them are seasonal. Population of the *haor*-based area is tending to increase every year. Despite the economic importance of the *haors*, people in the region are poorer than in any other part of the country due to geographical remoteness from the land.  In the *haor* areas, more than 28% of the total population lives below the Lower Poverty Line (LPL). Geographical remoteness is a constraint of development policy in the developing country and isolation from the growth center is also embarrassing the development plan.  Therefore, Sunamganj district is the mostly *haor*-based district in Bangladesh and purposively this *haor*-based district has been chosen for the study. In Sunamganj district there are 11 upzilas as stated earlier where one third of total area can be termed as wetlands. The *haor* basin of this district is an internationally important wetland ecosystem, which is situated in Dharmapasha upazila, Bishwamvarpur upazila, Derai upazila, Dowarabazar upazila, Jagannathpur upazila, Sullah upazila and Tahirpur upazila. These six upazilas has been selected for the study as they are known *haor*-based area nationally and internationally. Due to time and budget constraints the top 3 (three) upazilas out of these six has been selected as sample for the convenient in data collection. The three upazilas are. Bishwamvarpur upazila, Tahirpur upazila and Dharmapasha upazila. The study area is given in figure-3.1.

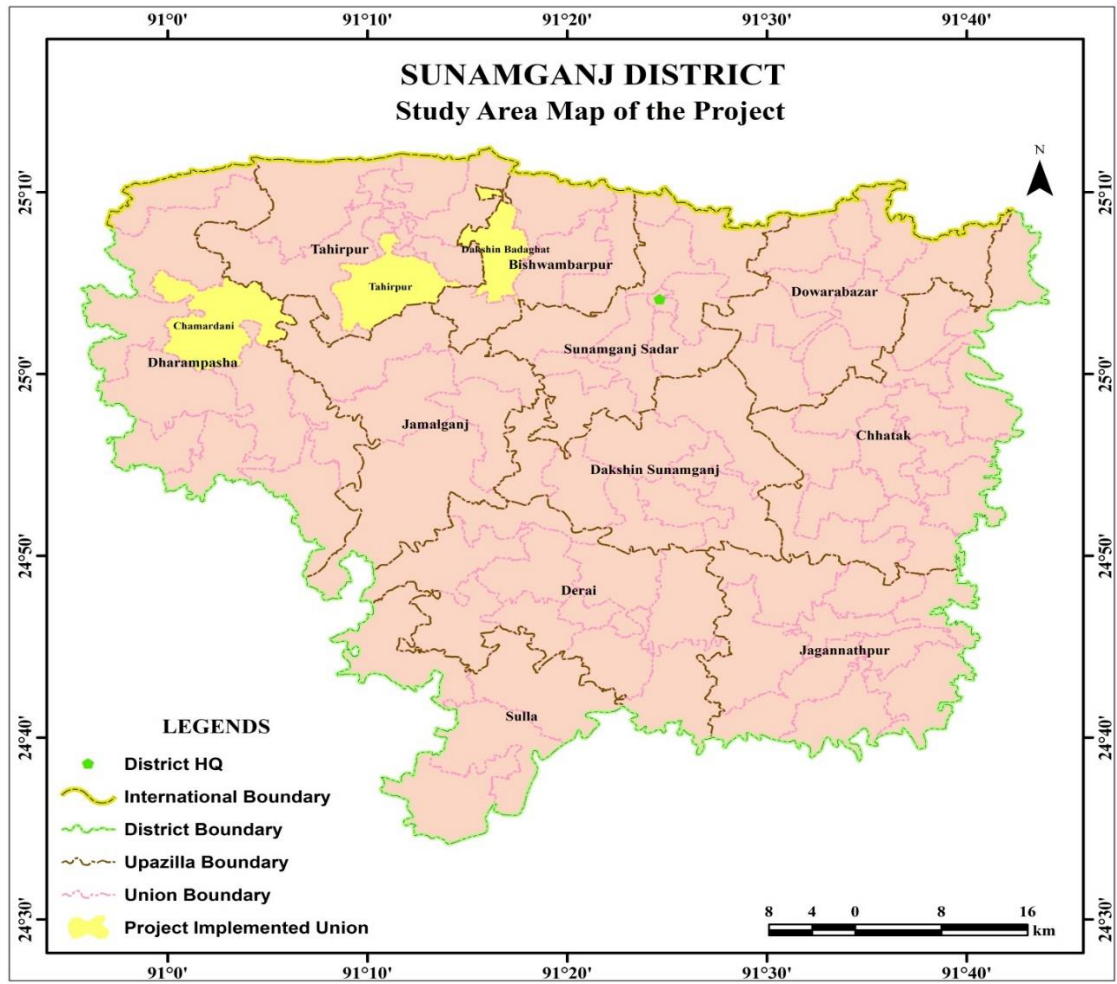

**Figure 3.1: Study Area.**

## 3.2 Sample and Sample Design

The study has gathered data from borrower of micro credit who had been in a microcredit program. Therefore, the borrower household (client) is the main unit of investigation. In measuring the impact of microcredit on clients, at first the clients have been divided into two groups on the basis of duration of microcredit inclusion. The first group was included those clients who are microcredit member for at least three years and the second group of respondents will consist of people who are not members of any microcredit program. In determining sample size, the following steps has been followed:

In the first stage, a sample has been selected from the study area under two categories-borrowers and non-borrowers. After that, a random sampling technique has been adopted to selected for 156 micro credit borrowers (members) from each upazila. Thus, the total sample size is 156×3 =468 micro credit borrowers have been selected for the interview process from the three upazilas. But in practice we have received 485 micro credit borrowers from survey.

In the second stage, we have selected 78 non-borrowers (control group) from each upazila randomly in our interview process. Thus, the second group (control group) is those people who had never been MC members until 78 agreed to participate from each upazila. Thus, total sample of control group is 78*3=234. But in practice we have received 228. The sample frame is given in the following table:

**Table 3.1: Sampling Frame of the Study.**

| Group of respondents | Upzila-1 | Upzila-2 | Upzila-3 | Total |
|---|---|---|---|---|
| Group 1: Members of microfinance | 160 | 162 | 163 | 485 |
| Group-2: Non-members of microfinance | 75 | 74 | 79 | 228 |
| Total sample size | 235 | 236 | 242 | 713 |

## 3.3 Questionnaire Design

A semi- structured questionnaire has been prepared for collecting data from the two groups of respondents as stated above in two parts. The first part is for group-1 respondents who are the microcredit members while the second part questionnaire for the group-2 who are not member of microcredit *i.e.*, control group. A qualitative survey, such as key-persons' interview and open group discussions has also been made for constructing, designing and finalizing questionnaire for the survey. Then a 'pre-test survey' has been conducted to finalize the questionnaire as well as instruction manuals for field supervisors and enumerators.

## 3.4 Collection of Data

The data from the respondents of all three groups *i.e.*, primary data has been collected through field survey by the data collector as the respondents not educated up to national level with the semi structured questionnaire (Annexure-Questionnaire). The members of micro-credit (clients) have been responded in two segments *viz.,* their socio-economic status before and after the member of MC *i.e.*, before and after borrowings. The secondary data has been collected from the related published and unpublished books, theses, reports and articles.

**3.5 Analysis of Data**

The Statistical Package for Social Sciences (SPSS) software has been used for the statistical analyses. The demographic and socioeconomic variable of the respondents have been coded and has been entered in computer using the SPSS software. Mainly two kinds of analyses have been made from the data base:

First-The descriptive statistics- The mean, standard deviation and percentage have been determined of the variables viz. the volume and duration of credit, respondent's annual income, annual expenditure, annual savings, knowledge about micro credit, expenditure or investment pattern of micro credit, purpose of loan causes of non- payment of loan and attitude of borrower's about micro-credit possession assets, possession of productive assets, savings, net worth, expenditure, financial capacity and demographic profile of all three groups as stated in the sample design.

Second- Difference in differences (DID) is used to measure the impact of micro credit on poverty alleviation. Under this method first we calculated the effect of a treatment i.e., an explanatory variable on an outcome (i.e., a response variable or dependent variable) by comparing the average change over time in the outcome variable for the treatment group to the average change over time for the control group.

Difference-in-differences (DID) Method

Difference in differences is typically used by project management to evaluate the impact of program intervention over a specific time gap. It is a quasi-experimental research design using observational data, by studying the differential effect of a treatment on a 'treatment group' versus a 'control group'. In reality, DID makes use of longitudinal data from treatment and control groups to obtain an appropriate counterfactual to estimate a causal effect.

To estimate the effect of benefits from micro-credit programs, this study employed DID to compare the changes in outcome variables over time between a household (HH) that is enrolled in the micro-credit program and a HHs that is not included (non-borrower) in the micro-credit program. With this method following steps have been followed to measure the impact of micro- credit by the HHs:

i.  Estimate the difference of the outcome variables related to socio-economic status by comparing the current status (at survey date 2019/20) of beneficiary HHs (Treatment group/borrowers) with their status before entering the program (the study considered before 3 years);

ii. Estimate the difference of the outcome variables related to socio-economic status by comparing the current status (at survey date) of non-beneficiary HHs (Control group/non-borrowers) with their status before 3 years of the survey;

iii. Estimation of the difference of the above two differences [(ii)-(i)] will yield the net impact of the benefits from micro-credit programs.

Factor Analysis

Factor analysis is a multivariate statistical technique that addresses itself to the study of interrelationships among a total set of observed variables (Manly, 2005; Rencher, 2002). The technique allows looking at groups of variables that tend to be correlated to one-another and identify underlying dimensions that explain these correlations. While in multiple regression model, one variable is explicitly considered as dependent variable and all the other variables as the predictors; in factor analysis all the variables are considered as dependent variables simultaneously. In a sense, each of the observed variables is considered as a dependent variable that is a function of some underlying, latent, and hypothetical set of factors. Conversely, one can look at each factor as dependent variable that is a function of the observed variables.

If $\{X_1, X_2, \ldots, X_n\}$ be a set of n observed variables and $\{F_1, F_2, \ldots, F_m\}$ be a set of unobservable variables then the factor analysis model can be expressed as

$$X - \mu = LF + \varepsilon$$

where $L_{n \times m}$ is the matrix of factor loadings (coefficient $l_{ij}$ is the loading of i-th variable on the j-th factor) and $\mu$ is a vector of the means of $X_i$.

Several methods are available in literature to estimate factor loadings and factor scores. The study considers principal component method to estimate the factor loadings and communalities [ $h_i^2 = \sum_{j=1}^{m} l_{ij}^2$ ], a measure of the variation of observed variables through factors. Several factor rotation methods like 'Varimax', 'Equamax', 'Quartimax' are adopted to find better estimates of factor loadings. Once the factors are identified and factor loading matrix is estimated then the estimated values of factors, factor scores, are calculated for each individual. The estimated values of factor scores are often used for diagnostic purposes as well as inputs to a subsequent analysis.

# CHAPTER FOUR
# ANALYSIS OF DATA

## 4.1 Demographic Profile of the Respondents

Appendix table 4.1 depicted that 66.2% respondents of borrowers and 98.7 non-borrowers are head of the family and an average 76.6% of both type of respondents. Among the borrower respondents 32% is husband/wife and 1.3% of non-borrowers and on average 22.2 of both type. It is clear that head and husband are highly    significantly to total.  In terms of sex 64.7% of borrowers and 92.5% of non-borrowers are male while 35.3% of borrowers and 7.5% of non-borrowers are female and in both cases of both are significantly related to sex ratio.  The age cat of respondents showed that 62.5% of borrowers and 66.7% of non–borrowers and in total 63.8% of both falls between 31 and 50.  The marital status of the respondents portrayed that   92.2% of borrowers and 92.5% of non –borrowers   and in total 92.3% are married.  The educational status of the respondents portrayed that    61.9% of borrowers and 61.8% of non –borrowers    and in total 61.9% are educated up to 5 years of schooling following no education 25.4% of borrowers and 18% of non-borrowers and 23% in total. The mean of schooling is $3.70 \pm 2.81$ years. The respondents by occupation showed that 24.1% of borrowers and 25.9% of non–borrowers   and in total 24.7% are day labor following household working 29.9% borrowers and 4.8% % of non –borrowers    and in total 21.9%, following farming 18.8% of  borrowers and 15.4% % of non –borrowers   and in total 17.7% .In terms of income earner the respondents showed that full time earner is 49.9%   of borrowers and  67.1 % % of non–borrowers   and in total 55.4%  following no work  32.2% borrowers and   17.1% % of non–borrowers    and in total 27.3%   and part time worker is 17.9% of borrowers and  15.8% non-borrowers and 17.3% in total.  The table showed that among the respondents 2.9% of borrowers and 3.5% of non- borrowers are disabling while 97.1% borrowers and 96.5% non-borrowers are able to work.

## 4.2 Demographic Profile of Household Members

The demographic profile of the household members of the respondents is shown in appendix table 4.2 in terms of different socioeconomic aspects.  The age structure of household members out of 3610 in total showed that 35.5%) falls between 0-15 following 28.80% falls between 16-30, 26% between 31-50 and 9.8% above 51. The age structure of the members of HHs is given in figure 4.2.1. The marital status of the members above age 16 showed that out of 2329 members 65.7% is married, 28.6% un- married and 5.7% is widows and separated. The analysis of educational statues of the members above age 7 portrayed that 52.1% has 1-5 years schooling following 19.9% 6-9 years, 15.7 with no education, 10.8% SSC and HSC and only 1.5% graduate and above. The education status of the members is given in figure 4.2.2.

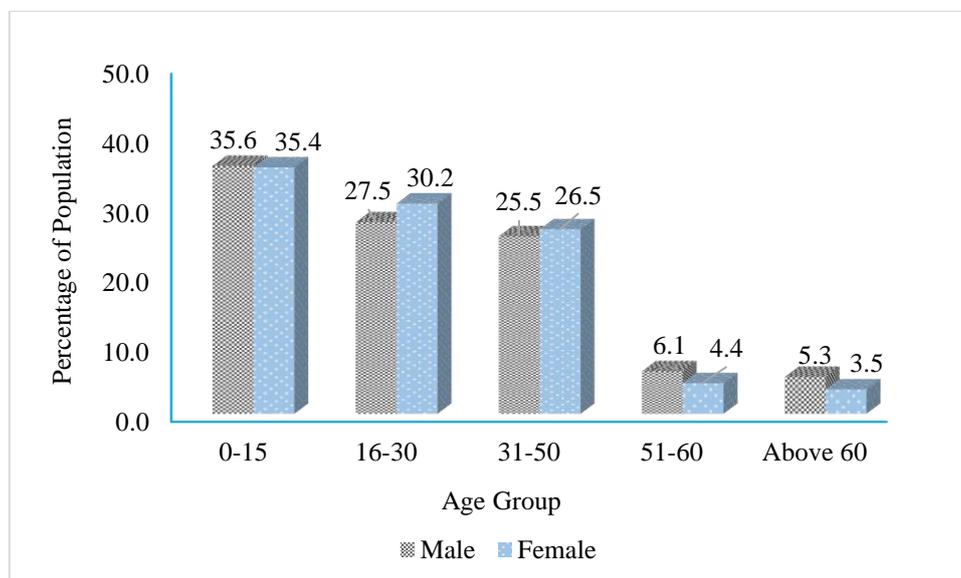

**Figure 4.2.1: Age Structure of the Members of HHs.**

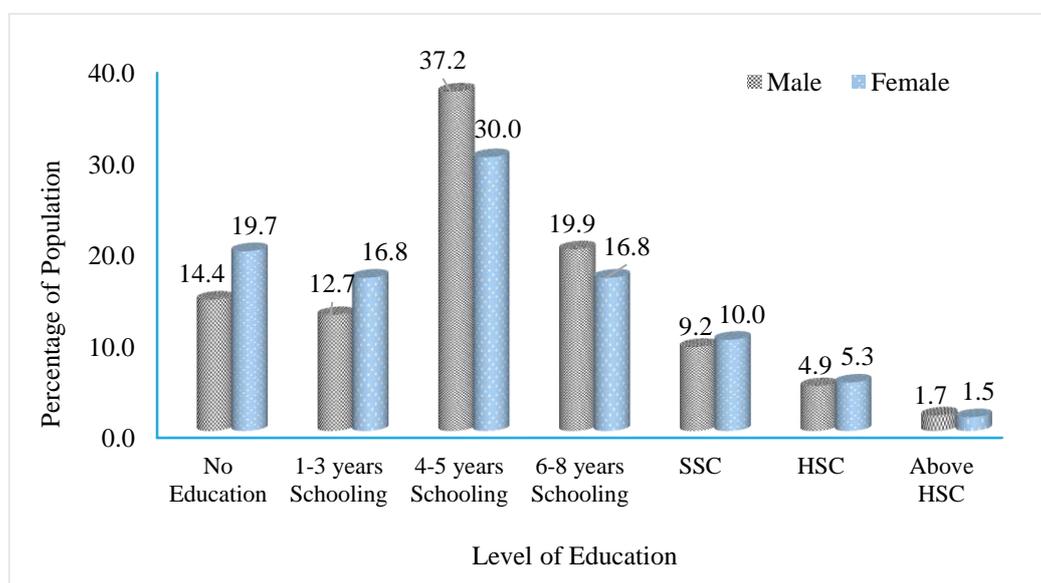

**Figure 4.2.2: Educational Status of the Members of HHs**

The occupation of the members of 16 to 60 aged stated that household works consisted of 32.3% following day labor 13.8%, student 12.6%, others 12.6%, service/business 10.8% off-farm activities 10.4% and farming 7.4%. The occupational status is given in figure 4.2.3.

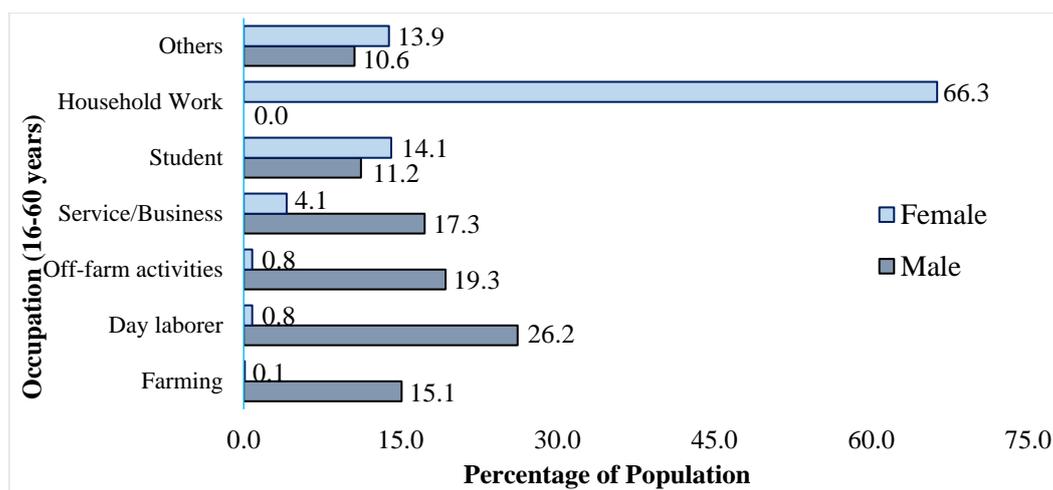

**Figure 4.2.3: Occupation of the Members of HHs.**

The income earner member above the age of 16 of the households showed that 57.3% members has no work while 30.30% full-time workers, 12.3% part-time workers and others 0.2%. It is observed that 4.5% member is disabling. The composition of households' members depicted that male female ratio is 100:93 while female headed is 2.6% and male headed is 97.4% and average family size is 5.6 o5 6 members per household. The dependency ratio is 55% in total where child (0–14) dependency ratio is 48.95% and aged (60+) dependency ratio is 6.05%. In depth analysis of the members of households in terms of education the researcher found the dropout and salient indications on education as given in table 4.2.1.

**Table 4.2.1: Dropouts and Basic Educational Indicators of the Household Population.**

| Indicators | Male | Female | Total |
|---|---|---|---|
| % of children (aged 6-11 years) not going to primary school | 9.9 | 10.9 | 10.4 |
| % of children (aged 12-18 years) not going to secondary school | 4.9 | 3.7 | 4.3 |
| % of adult members (aged 15 and above) never attended school | 15.7 | 22.5 | 19.0 |
| Average years of schooling of household head | 4.09 | 1.87 | 3.95 |
| Average years of schooling of wife/husband of household head | 2.25 | 3.35 | 3.32 |
| Average years of schooling of adult members aged 15 and above | 5.25 | 4.64 | 4.96 |
| **Total (n)** | **1880** | **1730** | **3610** |

Table 4.2.1 clarified that 10.4% children aged 6-11 years does not go to primary school while 4.3 children aged 12-18 years does not go to secondary school and 19% adult members aged above 15 never attended school. On the other hand, average schooling of household heads is 3.95 years; husband/wife for 3.32 years and adult members aged 15 and above is 4.96.

## 4.3 Housing Information of the Households

The housing information of the respondents is presented in appendix table 4.3 which tells that 97.1% of borrowers, 93% of non-borrowers and in total 95.8% of both owns housing ownership. The house size is measured in terms of number of room and the table shows that 35.1% owns three room following 28.1% two room, 22.2% four room and 14.7% single room while the average number of persons per room is 2.20. There is separate kitchen room in 60.7% houses. The houses are in four types as presented and the data showed that 51.2% holds tin shed roof and wall following 37% tin shed roof and muddy wall, 9.3% tin shed and *Pucca* wall and floor and 2.2% straw roof and bamboo/muddy wall. The findings on sources of cooking fuel showed that 48.4% households use wood/kerosin following 33.3% Straw/Leaf/Husk/Jute stick,16.1% cow dung,2.4% gas and 0.1%others. The sources of drinking water of the houses is89.8% from tube-well following 9.7 ponds and 0.6% supply water. Electricity coverage in the village or area of the study is 95.5% while the Electricity coverage in the house is 91.3% and the ownership of toilet by the households is 90.3%. Types of toilets used by household members are *Katcha* toilet 57.15 following *pucca* toilet (not water resistant) 35.3% *pucca* toilet (water resistant) 6.3% and open field/others is 1.3%.

## 4.4 Possession of Durable Assets by Households

The respondents are given nineteen general items and others as in table 4.4 and in figure 4 that are useable by the households and were asked to provide number of item(s) possesses by them. The data showed that among the top 5 assets 97.1 bed, following 96.8 households owns *Chauk* following, 91.7% mobile phone, 85.7% electric fan, 83.9% table/chair but only 19.5% holds television and 4.1% holds radio.

**Table 4.4: Possession of Durable Assets by Households Micro-credit Status.**

| Type of Assets | Type of Households | | | | Total | |
| --- | --- | --- | --- | --- | --- | --- |
| | Borrower | | Non-borrower | | | |
| | Number | % | Number | % | Number | % |
| Radio | 13 | 2.7 | 16 | 7.0 | 29 | 4.1 |
| Television | 71 | 14.6 | 68 | 29.8 | 139 | 19.5 |
| Mobile phone | 455 | 93.8 | 199 | 87.3 | 654 | 91.7 |
| Electric Fan | 408 | 84.1 | 203 | 89.0 | 611 | 85.7 |
| Almira | 215 | 44.3 | 109 | 47.8 | 324 | 45.4 |
| Cookeries | 355 | 73.2 | 209 | 91.7 | 564 | 79.1 |
| Cutleries | 355 | 73.2 | 209 | 91.7 | 564 | 79.1 |

| Table/Chair | 404 | 83.3 | 194 | 85.1 | 598 | 83.9 |
|---|---|---|---|---|---|---|
| Bed | 471 | 97.1 | 221 | 96.9 | 692 | 97.1 |
| *Chauki* | 468 | 96.5 | 222 | 97.4 | 690 | 96.8 |
| Drawing Room Furniture | 45 | 9.3 | 4 | 1.8 | 49 | 6.9 |
| Bicycle | 39 | 8.0 | 36 | 15.8 | 75 | 10.5 |
| Boat | 107 | 22.1 | 35 | 15.4 | 142 | 19.9 |
| Tube well | 82 | 16.9 | 54 | 23.7 | 136 | 19.1 |
| Deep Tube well | 3 | 0.6 | 0 | 0.0 | 3 | 0.4 |
| Watch | 197 | 40.6 | 65 | 28.5 | 262 | 36.7 |
| Shelf | 120 | 24.7 | 72 | 31.6 | 192 | 26.9 |
| *Alna* | 320 | 66.0 | 146 | 64.0 | 466 | 65.4 |
| Motorcycle | 5 | 1.0 | 0 | 0.0 | 5 | 0.7 |
| Others | 68 | 14.0 | 30 | 13.2 | 98 | 13.7 |
| **Total** | **485** | | **228** | | **713** | |

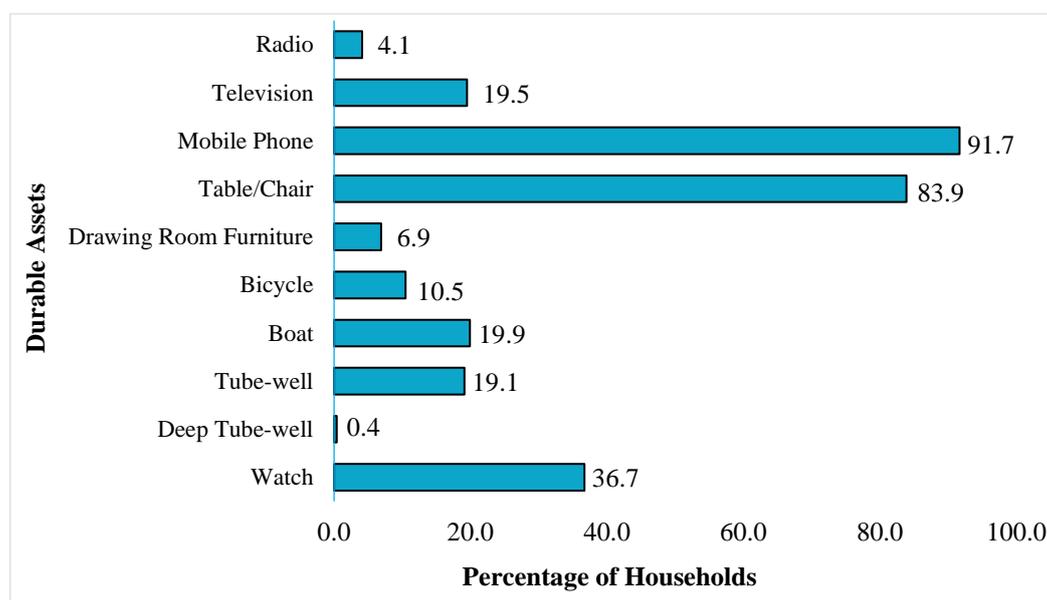

**Figure 4.4: Possession of Durable Assets.**

## 4.5 Possession of Productive Assets by Households

The productive assets are divided into two categories one is land and another is o non -land. These two are discussed below:

### 4.5.1 Landholdings by the households

The land holdings by the households are divided into: (i) homestead land; (ii) cultivable land (own); (iii) cultivable land (leased-in or sharecropped) and (iv) pond land. The details of the holdings of land are given in appendix table 4.5.1 and found that5.6% holds no homestead land, 79.5% holds 1 to 15 decimal, 7.4% holds 16-50 decimal and 7.4% holds more than 50 decimal. The data on cultivable land (own) explored that 71.1% landless following 15.4% holds more than 50 decimals, 6.7 holds 1 to 15 decimals also 6.7% holds    16-50 decimal. The holding status of cultivable Land (Leased-in or sharecropped) portrayed that 65.5% holds no land following 22.7% more than 50 decimal,7.4%16-50 decimal and 4.3% 1 to 15 decimals. The data on holdings of pond land showed that 96.4% has no pond land while 3.6% holds 1 to 15 Decimal. The landholding by HHs is given in figure 4.5.1:

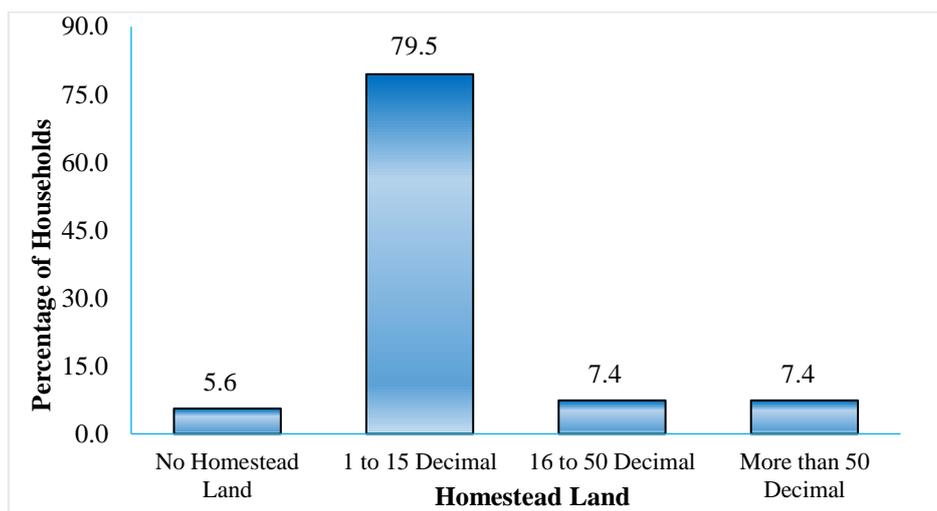

**Figure 4.5.1: Landholdings by Households**

### 4.5.2 Non-land productive asset of the households

There are eleven type non-land assets holds by the households as given in table 4.5.2. The data showed that197(41%) borrowers and 72(32%) non borrowers hold livestock following cultivation instruments by 161(33%) borrowers and 59(26%) non borrowers, fishing net 68(14%) borrowers and 43(28%) non borrowers. The combined   holdings by both borrowers and non- borrowers of these top three is 268(38%) 220(31%) and 111(16%) respectively.

**Table 4.5.2: Non-land productive assets of the households.**

| Type of Assets | Type of the Households | | | | Total | |
|---|---|---|---|---|---|---|
| | Borrower | | Non-borrower | | | |
| | N | Mean ± SD | N | Mean ± SD | N | Mean ± SD |
| Cultivation Instruments | 161 | 4.59 ± 2.68 | 59 | 3.71 ± 1.76 | 220 | 4.35 ± 2.49 |
| Livestock | 197 | 2.46 ± 1.62 | 72 | 2.60 ± 1.44 | 268 | 2.50 ± 1.57 |
| Rickshaw/Van | 13 | 1.62 ± 0.96 | 6 | 1.17 ± 0.41 | 19 | 1.47 ± 0.84 |
| Auto Rickshaw | 10 | 1.20 ± 0.63 | 9 | 1.00 ± 0.00 | 19 | 1.11 ± 0.46 |
| Sprayer | 2 | 1.50 ± 0.71 | - | - | 2 | 1.50 ± 0.71 |
| Fishing Net | 68 | 2.03 ± 1.01 | 43 | 2.00 ± 1.09 | 111 | 2.02 ± 1.04 |
| Bee Box | 5 | 1.0 ± 0.00 | - | - | 5 | 1.00 ± 0.00 |

| Sewing Machine | 13 | $1.15 \pm 0.38$ | 7 | $1.00 \pm 0.00$ | 20 | $1.10 \pm 0.31$ |
|---|---|---|---|---|---|---|
| Motor (Engine) | 13 | $1.15 \pm 0.38$ | 8 | $1.00 \pm 0.00$ | 21 | $1.10 \pm 0.30$ |
| Family Business | 40 | 71425.00 | 20 | 137250.00 | 60 | 93366.67 |
| Others | 16 | $6.69 \pm 9.36$ | 11 | $3.36 \pm 3.33$ | 27 | $5.33 \pm 7.59$ |
| **Total (n)** | | **485** | | **228** | | **713** |

### 4.6. Households' Knowledge about Micro-credit Benefit

In this section the knowledge about micro-credit benefits by the study households is discussed represented in appendix table 4.6. The table showed that 95% respondents have knowledge about micro-credit benefit while 85.6% tried to get those benefits. For getting credit firstly 60.2% attempted to nongovernment (MFI/NGO/Insurance) following 13.6% local Money Lender (Mohajan/Private Samittee and 6.9% to more than one sources and 1.5% to government (Banks/Co-operatives). After first attempt the households attempted in second time for getting credit and 9% tried to local Money Lender (Mohajan/Private Samittee following 3.9% Nongovernment (MFI/NGO/Insurance),2.8% Government (Banks/Co-operatives), 0.3% non-interest loan (Relatives/friends/neighbors) and only 0.1% to more than one sources. The respondents seek helped to get credit to different persons or bodies and among these 34.4% went to UP Office following 6.6% relatives/neighbors/friends, 5.5% NGOs, 3.4% government officer, 0.8% UP chairman and 0.7% to UP Member. Among the respondents 2.9% was asked to give money (bribe) for giving micro-credit benefit and 55% told that MC remove poverty.

### 4.7 Factors of Exclusion from Micro-credit Program

The researcher has identified some causes for which the non-borrowers who are qualified for micro-credit but not taken. There were fifteen causes in the schedule and the respondents were asked to on 5-point scaling as strongly disagree, disagree, neutral and agree and strongly agree. Appendix table 4.7 showed that among the causes the top five causes are strongly agreed by respondents are bureaucratic complexity (5%) following limitation of budget (3.8%), no political exposure (3.2%), didn't have any idea about such program and non-availability of collateral (2.2%). The top five causes agreed by the respondents are non-cooperation from local lenders (13%) following lack of networking or lobbying (11.4%), misappropriation of credit (11.2%), non-cooperation from public delegate of micro-credit institution (11.1%) and non-availability of collateral (9.4%). Through PCA among the fifteen reasons five factors are significantly mentionable for exclusion from micro credit based on Eigen value (more than 1.00) as seen in the Scree plot given below in figure 4.7

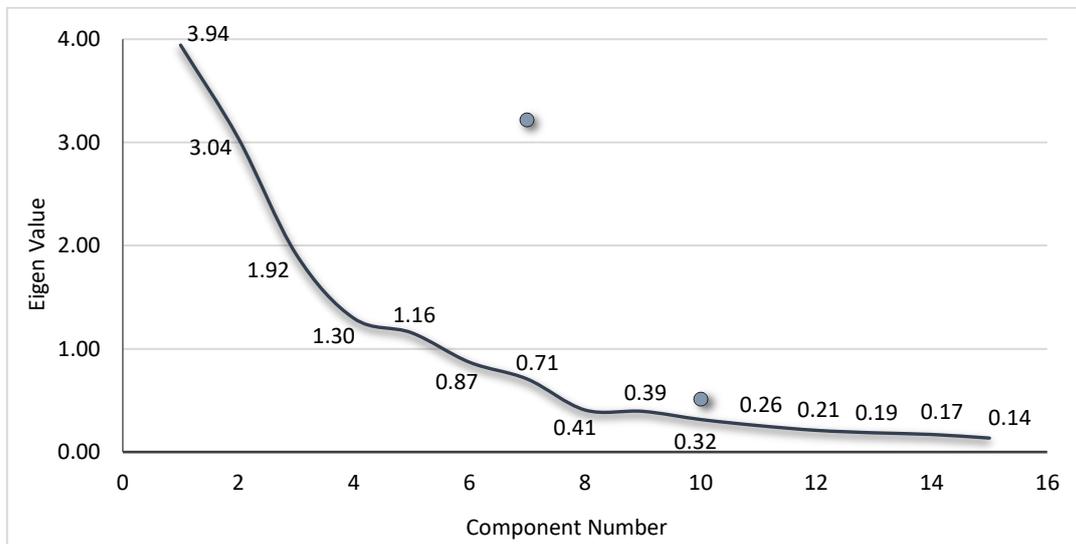

**Figure 4.7: Scree Plot of Factors of Exclusion from Micro-credit Program.**

The Principal Component Analysis (PCA) with Varimax Orthogonal rotation method explained 75.66% of total variation by the extracted factors with 0.703 Kaiser-Meyer-Olkin (KMO) measure of sampling adequacy. It means the sample size is sufficient to explore the factors. Details of the factors or causes are given in appendix table 4.7.1. The first component comprised of four reasons- (i) no micro-credit in the area; (ii) non availability of collateral; (iii) misappropriation of credit; and (iv) business. The second factor is consisted of four reasons- (i) non-cooperation from public delegate of micro-credit institution; (ii) Non-cooperation from local lenders; (iii) Lack of networking or lobbying and (iv) distance from micro-credit from the village. The third factor included three dimensions- (i) couldn't provide bribe or entry fee, (ii) limitation of budget (according to selector) and non-availability of NID. The fourth factor formed of two dimensions- (i) bureaucratic complexity and (ii) limitation of budget (according to selector). Lastly the fifth factor was made of two dimensions- (i) no political exposure and (ii) nepotism. The dimensional factors are given in table 4.7.1.

**Table 4.7.1: Major dimensional factors of excluded from the micro-credit benefits.**

| Serial | Reasons | Rotated Factor Loadings (Varimax) | | | | | Commu nalities |
|--------|---------|------|------|------|------|------|--------|
| | | F1 | F2 | F3 | F4 | F5 | |
| a | Bureaucratic complexity | | | | 0.848 | | 0.786 |
| b | Limitation of budget (according to selector) | | | | 0.810 | | 0.784 |
| c | Couldn't provide bribe or entry fee | | | 0.817 | | | 0.829 |
| d | No political exposure | | | | | 0.472 | 0.853 |
| e | Didn't have any idea about such program | | | 0.931 | | | 0.877 |
| f | Nepotism | | | | | 0.890 | 0.860 |
| g | Non-cooperation from public delegate of Micro-credit institution | | 0.511 | | | | 0.517 |
| h | Non-cooperation from local lenders | | 0.810 | | | | 0.711 |
| i | Non-availability of NID | | | 0.573 | | | 0.572 |
| j | Lack of networking or lobbying | | 0.742 | | | | 0.664 |
| k | Distance from Micro-credit from the village | | 0.644 | | | | 0.761 |
| l | No Micro-credit in the area | 0.775 | | | | | 0.798 |
| m | Non availability of collateral | 0.836 | | | | | 0.777 |
| n | Misappropriation of credit | 0.891 | | | | | 0.881 |
| o | Biasness | 0.763 | | | | | 0.681 |
| Percentage of Variation Explained | | 26.29 | 20.27 | 12.77 | 8.64 | 7.7 | |
| Total Variation explain by the extracted factors | | | | | | | 75.66 |
| Kaiser-Meyer-Olkin Measure of Sampling Adequacy | | | | | | | 0.703 |
| Bartlett's Test of Sphericity | | Chi-Square = 1688.22; df = 105 & P-value <0.001 | | | | | |
| Extraction Method | | Principal Component Analysis | | | | | |

## 4.8 Profile of Micro-credit Programs

The micro-credit programs available in the study area is investigated in terms of (i) inclusion month ;(ii) interest rate ;(iii) installment type;(iv) number total installment; (v) duration of loan in month; (vi) collateral or security;(vii) amount of paid loan and interest; and (vii) amount of unpaid loan and interest of both formal and informal loans. Appendix table 4.8 showed the data on profile of micro-credit. The data on month of inclusion depicted that both formal and informal loan are taken during October to December (133+59) following January to March (86+108), July to September (64+32) and April to June (75+14). The insight of interest rate depicted that there is 0% for 20 borrowers of informal credit only. The rate of interest varied between 1% to 25% and more for both of formal and informal credits. The borrowing rate 11% to 15% included maximum formal (143) and informal (160) borrowers following 21% to 25% formal (98) and informal (113), 16% to 20% formal (73) and informal (75), 1% to 10% formal (40) and informal (82) and more than 25% formal (4) and informal (35). The loans are paid in installment basis and the data showed that most of the loans of both formal (287) and informal (9) are on weekly basis following monthly formal (63) and informal (69) and biweekly, quarterly and annually are very small numbers while the number of installments are varied between i to more than 24 and the most of the loans of both formal and informal are in more than 24(formal 290,and informal 12,and total 302) installments following 12(formal 62and informal 43 total 105). It is remarkable that when the lion part of formal loans is paid in more than 24 installments then the informal loans are in single installment. The duration of loans is almost one year case of both formal and informal credits and paid against collateral. The payment of both types of loan is totally successful and both types are loans are paid regularly in time. All the loans are still going and remain a potion is unpaid.

### 4.9 Purpose of Loan

### 4.9.1 Descriptive on the major purposes of loan

The borrowers borrow money for different purposes and from pilot survey seventeen (17) purposes of loan are identified and the data is given in appendix table 4.9.1 showed that the top ten items of purposes in total of both formal and informal loans are purchasing food items is (32%) following crop production (26.2%), payment of previous loan (22.1%), trade/business/industry (20%), rearing cattle/poultry ranked (18.4%), healthcare expenditure (15.7%), repairing cost of houses (13.6%), constructing housing(8.7%) and fish farming/fishing (5.4%). The dimensional factor loading is given in appendix table 4.9.2.

### 4.9.2 Major Dimensional purpose of taking Loan

There are 17 purposes as stated in table 4.9.1 and these analysis by PCA with the are rotated to find out the major dimensions as in figure 4.9.2 and based on Eigen value (minimum 1.00 and above) six factors (purposes) are identified significant for taking loan. The scree plot is given below:

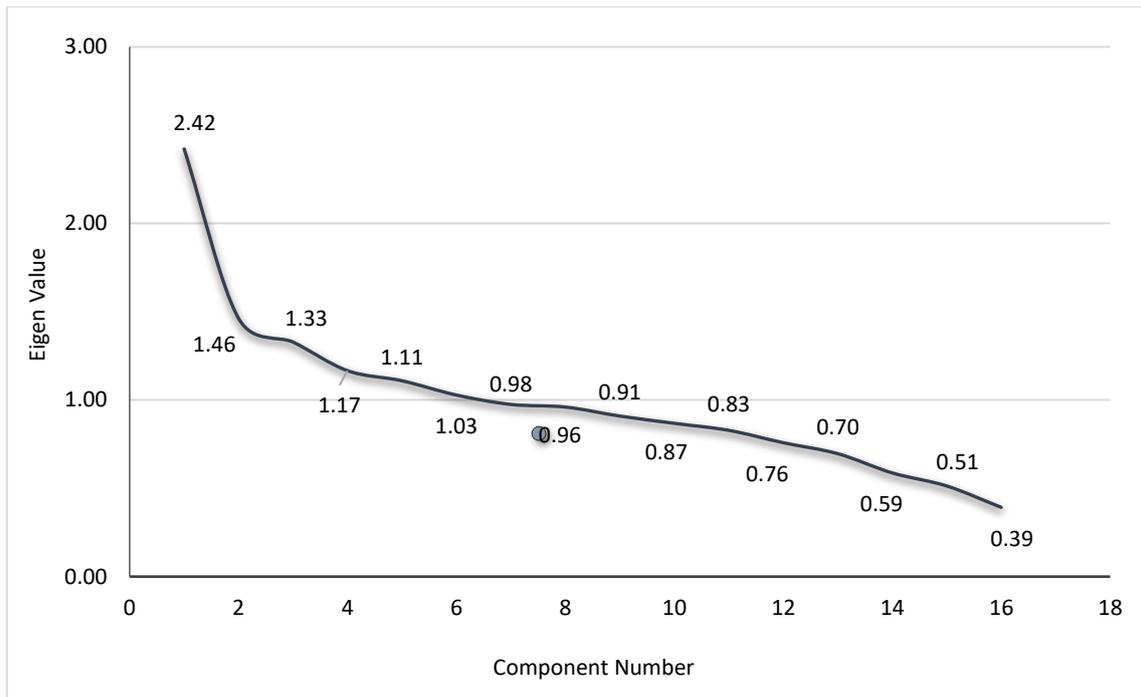

**Figure 4.9.2: Scree Plot of the Purpose of Taking Loan.**

The PCA with Varimax Orthogonal rotation method explained 53.22% of total variation by the extracted factors with 0.571 KMO measure of sampling adequacy. It means the sample size is sufficient to explore the purposes. Details of the factors or purposes are given in table 4.9.2 and the rotation of factor loading developed six dimensions. The first dimension explained 15.14% variations and composed of purchasing of food items, repairing cost of houses, healthcare expenditure and education. The second dimension explained 9.14% variations with the factors of rearing cattle/poultry, purchasing of livelihood and payment of loan equipment. The third dimension is consisted of crop production and tackling shocks of natural calamities with explaining 8.3% variations. The fourth dimension has explained 7.29% variations composing the tackling shocks of sudden death of HH head and others. The fifth dimension is composed of Sending family member to abroad fish farming/fishing by explaining 7.93% variations. The sixth dimension has explained 6.43% variations having trade/business/industry and daughter/son's marriage purposes. The factors in total explained 53.33% and Kaiser-Meyer-Olkin Measure of Sampling Adequacy is 0.57 which is closed to accept of 0.59.

**Table 4.9.2: Major dimensional purpose of taking loan**

| Serial | Reasons | Rotated Factor Loadings (Varimax) | | | | | | Communalities |
|---|---|---|---|---|---|---|---|---|
| | | F1 | F2 | F3 | F4 | F5 | F6 | |
| a | Purchasing of food items | 0.651 | | | | | | 0.476 |
| b | Crop production | | | 0.799 | | | | 0.714 |
| c | Rearing cattle/poultry | | -0.593 | | | | | 0.510 |
| d | Sending family member to abroad | | | | | 0.718 | | 0.629 |
| e | Trade/Business/Industry | | | | | | 0.399 | 0.676 |
| f | Fish farming/Fishing | | | | 0.596 | | | 0.478 |
| g | Daughter/son's marriage | | | | | | -0.836 | 0.740 |
| h | Constructing housing | | 0.558 | | | | | 0.524 |
| i | Tackling shocks of natural calamities | | | 0.643 | | | | 0.488 |
| j | Tackling shocks of sudden death of HH head | | | | -0.427 | | | 0.192 |
| k | Purchasing of livelihood equipment | | 0.494 | | | | | 0.404 |
| l | Payment of loan | | 0.481 | | | | | 0.445 |
| m | Repairing cost of houses | 0.578 | | | | | | 0.475 |
| n | Healthcare expenditure | 0.668 | | | | | | 0.477 |
| o | Education | 0.718 | | | | | | 0.544 |
| p | Others | | | | -0.826 | | | 0.746 |
| Percentage of Variation Explained | | 15.14 | 9.14 | 8.30 | 7.29 | 6.93 | 6.43 | |
| Total Variation explain by the extracted factors | | | | | | | | 53.222 |
| Kaiser-Meyer-Olkin Measure of Sampling Adequacy | | | | | | | | 0.571 |
| Bartlett's Test of Sphericity | | Chi-Square = 686.807; df = 120 & P-value <0.001 | | | | | | |
| Extraction Method | | Principal Component Analysis | | | | | | |

## 4.10 Expenditure and Investment Pattern of Loans

The respondents were asked to provide the usages of loans in different items in terms of amount and percentage of loans and for this reason the questionnaire was covered 24 heads of expenditure and investment out which fourteen heads were recognized as stated in table 4.10. The top seven (7) heads in total of both formal and informal credit are consumption on food (151) following purchasing agricultural inputs (102) payment of loan (93), family enterprises (89), purchasing animals (83) housing improvements (72) and health care (71).

**Table 4.10: Distribution of Loan among Different Heads of Expenditure in 2019/20.**

| Expenditure Heads | Type of Loan | | | | | |
|---|---|---|---|---|---|---|
| | Formal | | Informal | | Both | |
| | N | Average | N | Average | N | Average |
| Consumption on food | 117 | 11422.22 | 34 | 12235.29 | 151 | 11605.30 |
| Clothing and others essentials | 41 | 5012.20 | 10 | 11500.00 | 51 | 6284.31 |
| Purchasing agricultural inputs | 88 | 13734.09 | 14 | 13500.00 | 102 | 13701.96 |
| Purchasing durable goods | 13 | 8923.08 | 1 | 5000.00 | 14 | 8642.86 |
| Housing improvements | 62 | 10427.42 | 10 | 13950.00 | 72 | 10916.67 |
| Purchasing land | 13 | 13423.08 | 7 | 38000.00 | 20 | 22025.00 |
| Purchasing animals | 71 | 10753.52 | 12 | 12500.00 | 83 | 11006.02 |
| Payment of loan | 79 | 13373.42 | 14 | 14892.86 | 93 | 13602.15 |
| Family enterprises | 68 | 30875.00 | 21 | 24857.14 | 89 | 29455.06 |
| Health care | 54 | 7814.81 | 17 | 28705.88 | 71 | 12816.90 |
| Human capital | 11 | 29727.27 | 1 | 5000.00 | 12 | 27666.67 |
| Sending son to abroad | 7 | 55714.29 | - | - | 7 | 55714.29 |
| Daughter's/Son's marriage | 6 | 34666.67 | 3 | 31666.67 | 9 | 33666.67 |
| Others | 94 | 19819.15 | 55 | 21872.73 | 149 | 20577.18 |
| **Total (n)** | **358** | | **127** | | **713** | |

## 4.11 Socio-economic Impact of Micro-credit on borrowers:

The impact of micro credit is explained by comprising (i) total consumption expenditure, (ii) total investment expenditure (iii) savings; and (iv) total income of three years ago (2016/17) with in 2019/20 in terms of both formal and informal credit. Here is impact of informal credit is presented through table from 4.11.1 to table 4.11.7 are presented in the body and also in the appendices.

## 4.11.1 Impact of informal micro-credit through economic performance based on before-after comparison:

Appendix table 4.11.1 depicted that the annual average total income has been increased from Tk. 87393.7 before receiving loan to Tk. 103145.7 after receiving loan than that of before significantly and with this line labor sale has been increased significantly from Tk. 54737.4 to 68075.8 and all the components of total income showed an increase after receiving the loan except business. The annual average total expenditure has been increased from Tk. 90948.8 before receiving loan to Tk. 116220.5 after receiving loan than significantly. The consumption and investment are the main two element of expenditure and both have been increased significantly after credit than that of before. The food and non-food are the two sub-elements of consumption and both have been increased significantly after credit than that of before.

There are ten sub-elements of investment and among them agricultural, productive assets and others have been increased significantly. The savings which is the remains after deducting consumption from total income has been decreased insignificantly.

### 4.11.2 Impact of formal micro-credit through economic performance based on before-after comparison:

The sources of formal micro-credit are government and non-government banks, co-operatives, microcredit institutions and NGOs. The impact of credit from these sources is given in appendix table 4.11.2 which clarified that total income and its components – agricultural, non–agricultural and labor sale have been increased significantly though income from business and donation have been increased insignificantly. The amount of debt also increased without significance. The total expenditure and its two main components consumption and investment both have been increased significantly. Both the elements of consumption food and non- food has been increased significantly. The elements of investments - education & training, medical, agricultural, family business, household development and others have been increased significantly but savings increased insignificantly.

### 4.11.3 Impact of micro-credit in terms of formal and informal credit receiving households based on Difference-in-difference Method (DID)

Under DID method the impact of micro-credit is compared between treatment group (borrowers) and control group (non-borrowers). Appendix table 4.11.3 showed the over-all impact of both formal and informal credit. It portrayed that total income, total expenditure and investment have been increased 13.57%, 10.39% and 26.17%. All the elements of total income have been increased except debt which has been decreased by 2.39%. But the decrease in debt is the good sign of positive impact of debt. Consumption of food has been increased but non-food has been decreased. All the elements of investment have been increased except some factors. The savings has been decreased due excess increase in investment.

### 4.11.4 Percentage change of households' benefits using micro-credit based on did method

Appendix- table 4.11.4 showed that total income has been increased by 3.25% over the study period. All the elements of total income have been increased except labor sale while it has been decreased. It means that some of the borrowers have become self-reliance by using debt as debt has been also increased. The total expenditure has been increased by 8.76%. over the study period. The consumption expenditure is one of the elements of total expenditure has been decreased by 4.42% while investment (capital) expenditure has been increased at show high rate (31.3%) as a result saving has been decreased by 11.8%.

### 4.11.5 Percentage distribution of households' food insecurity during the study period

Table 4.11.5 showed that food security has been classified in terms of number of meals taken the respondents in a day and based on this criterion we divided the respondents into: (i)some periods of hunger (normal food insecurity) (iii) two meals a day throughout year (no food insecurity) and (iii) three meals a day throughout year (no food insecurity). The data showed that normal food insecurity of both borrowers and non-borrowers have been decreased during the study period and in total it has been decreased from 22.1% to 13% similarly no food insecurity of both borrowers and non- borrowers have been decreased during the study period and in total it has been decreased from 28.1% to 24.7% but finally no food insecurity has been increased of both borrowers and non-borrowers over the study period and in total it has been increased from 49.8% to 62.3%. Therefore, the use of micro credit ensures food security and reduces food insecurity.

**Table 4.11.5: Percentage distribution of households' food insecurity for the period of 2016/17 and 2019/20**

| Indicators | Household Type | | | | Total | |
| | Borrower | | Non-borrower | | | |
| | 2016/17 | 2019/20 | 2016/17 | 2019/20 | 2016/17 | 2019/20 |
|---|---|---|---|---|---|---|
| Some periods of hunger (Normal Food Insecurity) | 22.6 | 10.9 | 21.5 | 17.5 | 22.1 | 13.0 |
| Two meals a day throughout year (No Food Insecurity) | 31.0 | 27.4 | 19.7 | 18.9 | 28.1 | 24.7 |
| Three meals a day throughout year (No Food Insecurity) | 46.4 | 61.6 | 57.0 | 63.6 | 49.8 | 62.3 |

### 4.11.6. Percentage distribution of households self-assessed socio-economic status

Table 4.11.6 showed the role of micro -credit graduation of households from lower class to upper class.  For this purpose, the households have been divided into (i) Extremely poor (ii) Moderately Poor (iii) Poor; (iv) Middle -class and (v) Rich. The analysis of data  found that both borrowers and non-borrowers under extremely poor has been decreased over the study period and in total it has been decreased from 25.5% to 13.9% similarly both borrowers and non-borrowers under moderately poor has been decreased over the study period and in total it has been decreased from 31.7% to 26.6% while both borrowers and non-borrowers under poor has been increased over the study period and in total it has been increased from 36.6% to 50.4% and both borrowers and non-borrowers under middle class has been increased over the study period and in total it has been increased from 4.9% to 7.9% and yet there is no change in rich category. So, microcredit helps to come out from extremely poor to moderately poor and from to moderately poor to poor and from moderately poor to poor be middle class and in long run it will help to be rich from middle class.

**Table 4.11.6: Percentage distribution of households self-assessed socio-economic status for the period of 2016/17 and 2019/20**

| Indicators | Household Type | | | | Total | |
|---|---|---|---|---|---|---|
| | Borrower | | Non-borrower | | | |
| | 2016/17 | 2019/20 | 2016/17 | 2019/20 | 2016/17 | 2019/20 |
| Extremely poor | 27.4 | 13.8 | 21.5 | 14.0 | 25.5 | 13.9 |
| Moderately Poor | 37.5 | 29.1 | 19.3 | 21.5 | 31.7 | 26.6 |
| Poor | 32.8 | 52.0 | 44.7 | 46.9 | 36.6 | 50.4 |
| Middle class | 2.3 | 5.2 | 10.5 | 13.6 | 4.9 | 7.9 |
| Rich | 0.0 | 0.0 | 3.9 | 3.9 | 1.3 | 1.3 |

## 4.11.7. Impact of micro-credit on changing educational and healthcare expenditure in 2019/20 compared to 2016/17

Table4.11.7 represented that education expenditure has been increased in case of both borrowers and non –borrowers over the study period and in total it has been increased from 18.4% to 58.8% while no change has been decreased of both borrowers and non –borrowers and in total it has been decreased from 55.5% to 38%. in the same time of both borrowers and non–borrowers over the study period have been decreased which in total from 26.1% to 3.2%. Table4.11.7 also represented that health expenditure has been increased in case of both borrowers and non–borrowers over the study period and in total it has been increased from 29% to 57.6% while no change has been decreased of both borrowers and non –borrowers and in total it has been decreased from 49.2% to 33.5%. in the same time of both borrowers and non–borrowers over the study period have been decreased which in total from 21.7% to 8.8%. Therefore, micro-credit has positive impact on changing on both education and health development.

**Table 4.11.7: Perceptions on change on educational and healthcare expenditure in 2019 compared to 2016.**

| Indicators | Household Type | | | | Total | |
|---|---|---|---|---|---|---|
| | Borrower | | Non-borrower | | | |
| | 2016/17 | 2019/20 | 2016/17 | 2019/20 | 2016/17 | 2019/20 |
| **Status of Education Expenditure** | | | | | | |
| Increased | 18.6 | 64.7 | 18.0 | 46.1 | 18.4 | 58.8 |
| No Change | 54.8 | 32.2 | 57.0 | 50.4 | 55.5 | 38.0 |
| Decreased | 26.6 | 3.1 | 25.0 | 3.5 | 26.1 | 3.2 |
| **Status of Health Expenditure** | | | | | | |
| Increased | 31.3 | 61.2 | 24.1 | 50.0 | 29.0 | 57.6 |
| No Change | 47.0 | 29.9 | 53.9 | 41.2 | 49.2 | 33.5 |
| Decreased | 21.6 | 8.9 | 21.9 | 8.8 | 21.7 | 8.8 |

## 4.12 Causes of non-payment of loan

To identify the factors responsible for non-payment of loan we have listed 15 factors measured in term of percentage of the respondents. The collected data is analyzed and presented in table 12.1 as below:

Table 12.1 portrayed that among the 15 factors rate of interest is very high is agreed by 64.2% of formal borrowers following 60.1% of installment period is very short, 50.8% of investment loss 47.8% of acute food problem and natural calamities 46.9%. While in case of formal credit 84.3% of rate of interest is very high following 77.2 of misappropriation of loan, 75.6% of Installment period is very short, 72.4% of both acute food problem and investment loss and 63.8% of natural calamities. In total, rate of interest is very high stood (69.5%), installment period is very short is in second (64.1%) investment loss is in third (56.5%), acute food problem is in forth (54.2%), and medical treatment/medicine is in the fifth (49.5). Therefore, in case of both formal and in-formal cases the top most five factors of non-payment of loan are about same and they are rate of interest is very high, installment period is very short, investment, acute food problem, and natural calamities.

**Table 12.1: Causes of non-payment of loan in time.**

| Name of the reasons | Type of Micro-credit | | | | | | | | Overall |
|---|---|---|---|---|---|---|---|---|---|
| | Formal | | | | Informal | | | | |
| | N | Disagree | Neutral | Agree | N | Disagree | Neutral | Agree | Agree |
| Acute food problem | 266 | 8.7 | 17.9 | 47.8 | 109 | 10.2 | 3.1 | 72.4 | 54.2 |
| Medical treatment/medicine | 265 | 3.4 | 26.5 | 44.1 | 109 | 6.3 | 15.0 | 64.6 | 49.5 |
| Investment loss | 270 | 5.9 | 18.7 | 50.8 | 109 | 6.3 | 7.1 | 72.4 | 56.5 |

| Natural calamities | 270 | 10.6 | 17.9 | 46.9 | 109 | 19.7 | 2.4 | 63.8 | 51.3 |
|---|---|---|---|---|---|---|---|---|---|
| Insufficient of loan for investment | 264 | 7.0 | 22.1 | 44.7 | 109 | 5.5 | 27.6 | 52.8 | 46.8 |
| Duration of loan is short for return from investment | 264 | 2.2 | 23.5 | 48.0 | 109 | 3.9 | 29.1 | 52.8 | 49.3 |
| Installment period is very short | 264 | 1.7 | 12.0 | 60.1 | 109 | 4.7 | 5.5 | 75.6 | 64.1 |
| Rate of interest is very high | 267 | 2.5 | 7.8 | 64.2 | 109 | 0.8 | 0.8 | 84.3 | 69.5 |
| Renewal of loan is unavailable | 264 | 17.0 | 33.0 | 23.7 | 109 | 20.5 | 36.2 | 29.1 | 25.2 |
| Misappropriation of loan | 265 | 7.5 | 24.9 | 41.6 | 109 | 4.7 | 3.9 | 77.2 | 50.9 |
| Crop's failure | 266 | 11.7 | 33.0 | 29.6 | 109 | 22.0 | 15.0 | 48.8 | 34.6 |
| Expenditure for marriage of son/daughter etc. | 264 | 32.1 | 30.7 | 10.9 | 109 | 70.9 | 4.7 | 10.2 | 10.7 |
| Family legal problems and expenditure | 264 | 36.0 | 30.0 | 7.5 | 108 | 71.7 | 6.3 | 7.1 | 7.4 |
| Unexpected accident | 264 | 27.4 | 31.6 | 14.8 | 108 | 74.0 | 5.5 | 5.5 | 12.4 |
| **Total (n)** | **358** | | | | **127** | | | | |

The Scree plot based on Eigen value 1.00 and above has identified four causes as significant as given in figure 4.12. The scree plot of causes of non-payment of loan is given below in figure 4.12. The compositions of the causes are given in table 12.2 by PCA analysis.

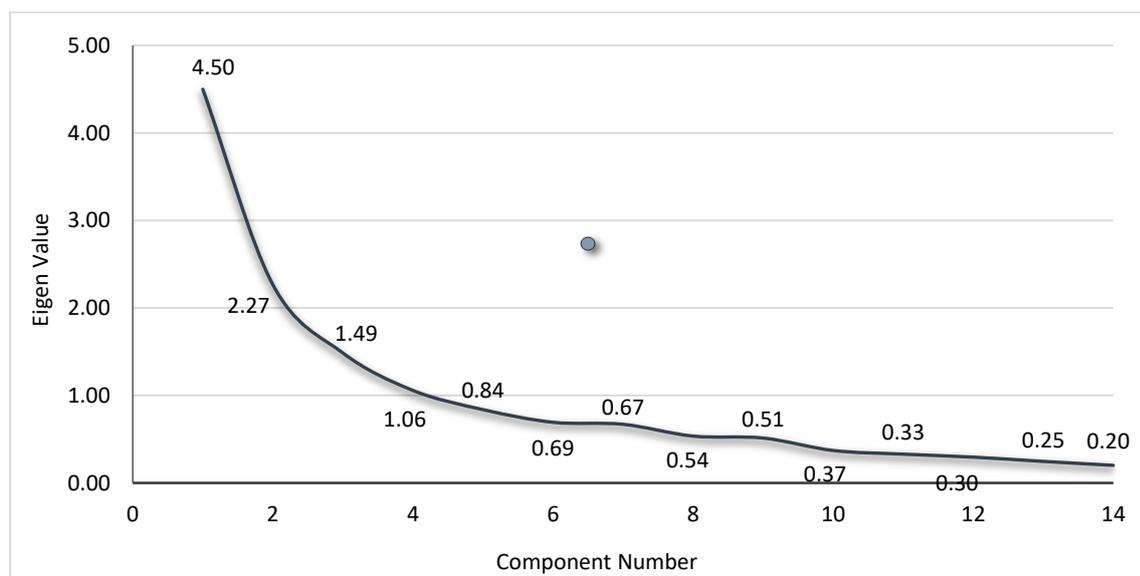

**Figure 4.12: Causes of Non-payment of Loan.**

Through the PCA with Varimax Orthogonal rotation method explained 66.48 % of total variation by the extracted factors with 0.791 KMO measure of sampling adequacy. It means the sample size is sufficient to explore the factors. Details of the factors or causes are given in table 4.12.2. The first factor is composed of six factors with 32.14(around 50% of 66.48%) percentage of total variation explained and they are (i) acute food problem; (ii) insufficient of loan for investment;(iii) duration of loan is short for return from investment; (iv) installment period is very short; (v) rate of interest is very high and (vi) renewal of loan is unavailable. The second factor is composed of three factors with 16.19 (around 25% of 66.48%) percentage of total variation explained and they are (i) expenditure for marriage of son/daughter; (ii) family legal problems and expenditure and (iii) unexpected accident. The third factor is composed of three factors with 10.61 (around 15% of 66.48%) percentage of total variation explained (i) medical treatment/medicine; (ii) natural calamities and (iii) crop's failure lastly the fourth factor composed of two factors with 7.55 (more than 10% of 66.48%) percentage of total variation explained and they are investment loss and misappropriation of loan.

**Table 4.12.2: Major dimensional causes of non-payment of loan timely**

| Serial | Reasons | Rotated Factor Loadings (Varimax) | | | | Commu nalities |
|---|---|---|---|---|---|---|
| | | F1 | F2 | F3 | F4 | |
| a | Acute food problem | 0.586 | | | | 0.476 |
| b | Medical treatment/medicine | | | 0.484 | | 0.714 |
| c | Investment loss | | | | 0.867 | 0.510 |
| d | Natural calamities | | | 0.823 | | 0.629 |
| e | Insufficient of loan for investment | 0.718 | | | | 0.676 |
| f | Duration of loan is short for return from investment | 0.854 | | | | 0.478 |
| g | Installment period is very short | 0.705 | | | | 0.740 |
| h | Rate of interest is very high | 0.578 | | | | 0.524 |
| i | Renewal of loan is unavailable | 0.427 | | | | 0.488 |
| j | Misappropriation of loan | | | | 0.471 | 0.192 |
| k | Crop's failure | | | 0.715 | | 0.404 |
| l | Expenditure for marriage of son/daughter etc. | | 0.837 | | | 0.445 |
| m | Family legal problems and expenditure | | 0.878 | | | 0.475 |
| n | Unexpected accident | | 0.890 | | | 0.477 |
| Percentage of Variation Explained | | 32.14 | 16.190 | 10.61 | 7.55 | |
| Total Variation explain by the extracted factors | | | | | | 66.480 |
| Kaiser-Meyer-Olkin Measure of Sampling Adequacy | | | | | | 0.791 |
| Bartlett's Test of Sphericity | | Chi-Square = 2157.331; df = 91 & P-value <0.001 | | | | |
| Extraction Method | | Principal Component Analysis | | | | |

**4.13 Attitude of Borrowers on Micro-credit**

To measure the attitude of respondents towards micro-credit we have listed 16 factors measured in term of percentage of the respondents with respect to disagree, neutral and agree. The collected data is analyzed and presented in appendix table 4.13. Appendix table 4.13 portrayed that among the 16 factors in case of formal credit among the top five factors showed that the rate of interest of micro-credit is reasonable is disagreed by 74.6% and agreed only by 17% following duration of credit is sufficient disagreed by 61.7% and agreed only by 21.2%, amount of credit is sufficient disagreed by 47.5% and agreed only by 35.5 %, by micro-finance your savings has increased is disagreed by 37.4% and agreed only by 32.4% and terms and conditions are not rigid disagreed by 30.4% and agreed only by 57.5%. Side by side in case of informal credit table 4.13 portrayed that among the 16 factors the top five factors showed that the rate of interest of micro-credit is reasonable is disagreed by 94.5% and agreed only by 3.1% following duration of credit is sufficient disagreed by 78% and agreed only by 18.9%, amount of credit is sufficient disagreed by 77.2% and agreed only by 19.7%, by micro-finance your savings has increased is disagreed by 66.9% and agreed only by 9.4 % and terms and conditions are not rigid disagreed by 57.5% and agreed only by 37.5%. Over-all we found that the top five factors are the rate of interest of micro-credit is reasonable is disagreed by 79.8% and agreed only by 13.4% following duration of credit is sufficient disagreed by 66.0% and agreed only by 20.6%, amount of credit is sufficient disagreed by 55.3% and agreed only by 31.3%, by micro-finance your savings has increased is disagreed by 45.2% and agreed only by 26.4 % and terms and conditions are not rigid disagreed by 37.5% and agreed only by 30.9%. Therefore, there are very many similarities of attitude towards micro-credit of the borrowers in terms of formal credit, informal credit and in over-all of the two. Through the PCA with Varimax Orthogonal rotation method explained 60.885 % of total variation by the extracted factors with 0.768 Kaiser-Meyer-Olkin (KMO) measure of sampling adequacy. It means the sample size is sufficient to explore the factors. The PCA analysis showed that four attitudes are significant based on Eigen value (1.00 and above) as seen in figure 4.13. The scree plot of major dimensional attitude of borrowers on micro-credit is given in figure 4.13

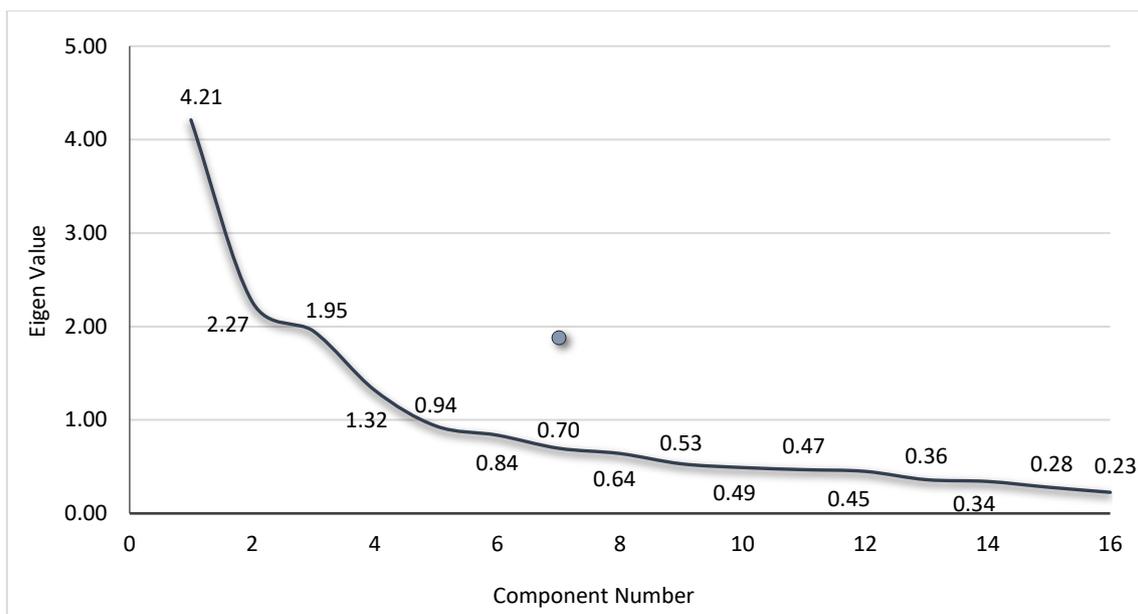

**Figure 4.13: The Scree Plot of Major Dimensional Attitude of Borrowers on Micro-credit.**

The PCA as presented in table 4.13.1 showed that through the Varimax rotated factor loading four factors are identified with 60.885% total variation explained after factors are extracted. The first factor has explained 26.316% of variations with six factors and they are (i) amount of credit is sufficient; (ii) by micro-finance your income has increased; (iii) by micro-finance your savings has increased; (iv) micro-finance is helping you in better access to education; (v) micro-finance is helping you in better access to healthcare; and (vi) due to micro-finance, employment opportunities have been increased. The second factor with explaining 14.154% of variation is composed of four sub factors which are (i) local loans are easier to get than MFIs; (ii) local lenders are friendly than MFIs; (iii) cost of local loans is lower than MFIs; and (iv) terms and conditions of local loans are easier than MFIs. The third principal factor with explaining 12.185% of variation is composed of three factors which are (i) rate of interest of micro-credit is reasonable; (ii) duration of credit is sufficient; and (iii) by micro-finance your food security has increased. The fourth principal factor with explaining 8.230% of variation is composed of three factors which are (i) terms and conditions are not rigid; (ii) micro-finance is helping you in better financial situation of your; and (iii) operational assistance received from MFIs was helpful to run the business. The KMO value is 0.768 which indicates that the sample is sufficient for analysis.

**Table 4.13.1: Major dimensional attitude of borrowers on micro-credit**

| Serial | Reasons | Rotated Factor Loadings (Varimax) | | | | Commu nalities |
|---|---|---|---|---|---|---|
| | | F1 | F2 | F3 | F4 | |
| 01. | The rate of interest of micro-credit is reasonable | | | 0.758 | | 0.576 |
| 02. | Amount of credit is sufficient | 0.482 | | | | 0.448 |
| 03. | Duration of credit is sufficient | | | 0.833 | | 0.704 |
| 04. | Terms and conditions are not rigid | | | | 0.472 | 0.553 |
| 05. | By micro-finance your food security has increased | | | 0.489 | | 0.449 |
| 06. | By micro-finance your income has increased | 0.760 | | | | 0.645 |
| 07. | By micro-finance your savings has increased | 0.794 | | | | 0.675 |
| 08. | Micro-finance is helping you in better access to education | 0.755 | | | | 0.674 |
| 09. | Micro-finance is helping you in better access to healthcare | 0.671 | | | | 0.545 |
| 10. | Micro-finance is helping you in better financial situation of your family | | | | 0.791 | 0.693 |
| 11. | Operational assistance received from MFIs was helpful to run the business | | | | 0.782 | 0.614 |
| 12. | Due to micro-finance, employment opportunities have been increased | 0.718 | | | | 0.528 |
| 13. | Local loans are easier to get than MFIs | | 0.584 | | | 0.609 |
| 14. | Local lenders are friendly than MFIs | | 0.836 | | | 0.813 |
| 15. | Cost of local loans is lower than MFIs | | 0.679 | | | 0.502 |
| 16. | Terms and conditions of local loans are easier than MFIs | | 0.824 | | | 0.713 |
| Percentage of Variation Explained | | 26.316 | 14.154 | 12.185 | 8.230 | |
| Total Variation explain by the extracted factors | | | | | | 60.885 |
| Kaiser-Meyer-Olkin Measure of Sampling Adequacy | | | | | | 0.768 |
| Bartlett's Test of Sphericity | | Chi-Square=2679.734; df=120 & P-value<0.001 | | | | |
| Extraction Method | | Principal Component Analysis | | | | |

## 4.14 Causes of Not Over-coming from Vicious Cycle of Poverty

As earlier we found in table 4.11.6 that there is a large number of households remain extremely poor, poor and moderately poor. It means a portion of households has fallen in vicious cycle of poverty because they are poor. In table 4.14.1 the causes of not overcoming from vicious cycle of poverty is described. Thirteen causes are listed and the respondents marked on three-point scales which are agree, neutral and disagree. It is found that the most five causes agreed by non-borrowers are Natural calamity (89%), following not getting loan in time (77.2) loss of investment (75.9), high interest (75.4%) and number of dependent members are high (69.7%). In the same way the most five causes agreed by borrowers are high interest (89.1%) following pressure of loan payment (76.5%), duration of loan is insufficient (71.5%), Insufficient loan (59%), Natural calamity (58.8%) and loss of investment (58.1%). In total the top five agreed causes of not over-coming from vicious cycle of poverty are high interest (84.7%) following pressure of loan payment (71.2%), natural calamity (68.4%), duration of loan is insufficient (64.7%) and loss of investment (63.8%).

**Table 4.14.1 Descriptive of the Causes of Not Over-coming from Vicious Cycle of Poverty.**

| Statements | Type of Households | | | | | | Overall | |
|---|---|---|---|---|---|---|---|---|
| | Non-borrower | | | Borrower | | | | |
| | Disagree | Neutral | Agree | Disagree | Neutral | Agree | Disagree | Agree |
| Insufficient loan | 17.1 | 36.0 | 46.9 | 17.7 | 23.3 | 59.0 | 17.5 | 55.1 |
| Duration of loan is insufficient | 10.5 | 39.5 | 50.0 | 16.3 | 12.2 | 71.5 | 14.4 | 64.7 |
| High interest | 1.3 | 23.2 | 75.4 | 5.6 | 5.4 | 89.1 | 4.2 | 84.7 |
| Renewal of loan not get | 12.3 | 61.4 | 26.3 | 30.7 | 42.7 | 26.6 | 24.8 | 26.5 |
| Diversion of loan | 1.3 | 71.9 | 26.8 | 8.9 | 52.6 | 38.6 | 5.6 | 34.8 |
| Loss of investment | 5.3 | 18.9 | 75.9 | 6.2 | 35.7 | 58.1 | 5.9 | 63.8 |
| Natural calamity | 1.8 | 9.2 | 89.0 | 5.8 | 35.5 | 58.8 | 4.5 | 68.4 |
| Not getting loan in time | 2.6 | 20.2 | 77.2 | 23.7 | 22.1 | 54.2 | 17.0 | 61.6 |
| Pressure of loan payment | 5.7 | 34.2 | 60.1 | 5.2 | 18.4 | 76.5 | 5.3 | 71.2 |
| Number of dependent members are high | 4.8 | 25.4 | 69.7 | 7.0 | 36.3 | 56.7 | 6.3 | 60.9 |
| Income earning member is absent | 27.2 | 19.3 | 53.5 | 30.7 | 27.8 | 41.4 | 29.6 | 45.3 |
| Number of dependent incomes earning person is low | 8.3 | 30.7 | 61.0 | 8.2 | 43.1 | 48.7 | 8.3 | 52.6 |
| I've not taken loan | 21.1 | 14.5 | 64.5 | 74.6 | 20.2 | 5.2 | 57.5 | 24.1 |
| **Total (n)** | **228** | | | **485** | | | **713** | |

Through the PCA with Varimax Orthogonal rotation method explained 58.63% of total variation by the extracted factors with 0.645 KMO measure of sampling adequacy. It means the sample size is sufficient to explore the factors. The PCA analysis showed that four attitudes are significant based on Eigen value (1.00 and above) as seen in figure 4.14.1

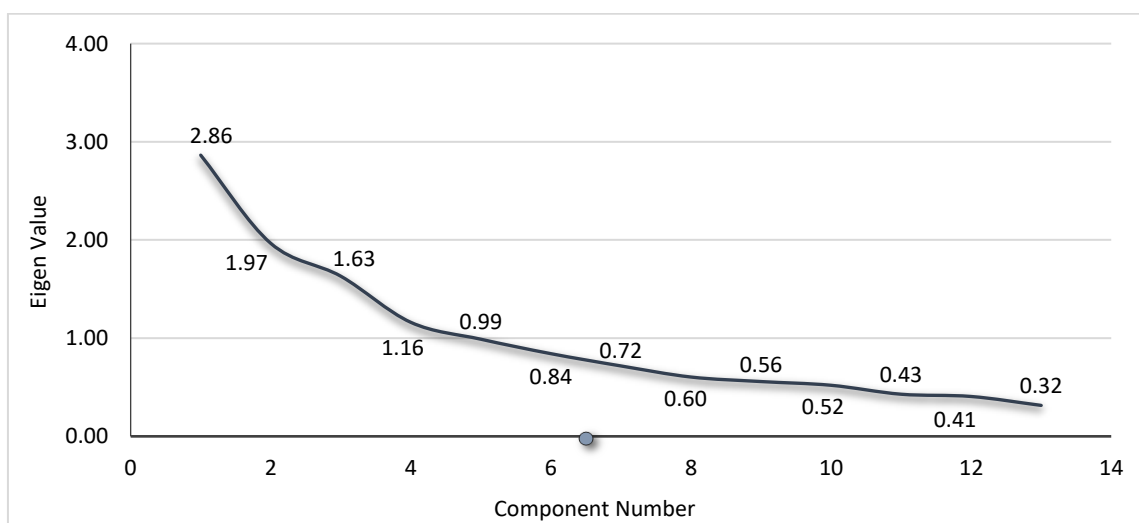

**Figure 4.14.1: The Scree Plot of Major Dimensional Causes of Not Over-coming from Vicious Cycle of Poverty.**

Table 4.14.2 showed that four Principal causes extracted from 13 causes with explaining around 59% percent variations. The first factor is composed with five components as natural calamity, not getting loan in time, number of dependent members is high, income earning member is absent and number of dependent incomes earning person is low. The second factor is composed with three components and they are insufficient loan, Duration of loan is insufficient and Renewal of loan not get. The third factor includes two components as diversion of loan and loss of investment. In the same time there are three components under fourth factor which are high interest, pressure of loan payment, I've not taken loan.

**Table 4.14.2: Major dimensional causes of not over-coming from vicious cycle of poverty**

| Serial | Reasons | Rotated Factor Loadings (Varimax) | | | | Commun alities |
|---|---|---|---|---|---|---|
| | | F1 | F2 | F3 | F4 | |
| 01. | Insufficient loan | | 0.634 | | | 0.433 |
| 02. | Duration of loan is insufficient | | 0.766 | | | 0.691 |
| 03. | High interest | | | | 0.755 | 0.655 |
| 04. | Renewal of loan not get | | 0.795 | | | 0.679 |
| 05. | Diversion of loan | | | 0.705 | | 0.604 |
| 06. | Loss of investment | | | 0.786 | | 0.669 |
| 07. | Natural calamity | 0.558 | | | | 0.535 |
| 08. | Not getting loan in time | 0.451 | | | | 0.432 |
| 09. | Pressure of loan payment | | | | 0.813 | 0.705 |
| 10. | Number of dependent members are high | 0.707 | | | | 0.509 |
| 11. | Income earning member is absent | 0.690 | | | | 0.522 |
| 12. | Number of dependent incomes earning person is low | 0.775 | | | | 0.614 |
| 13. | I've not taken loan | | | | -0.475 | 0.574 |
| Percentage of Variation Explained | | 22.031 | 15.126 | 12.544 | 8.933 | |
| Total Variation explain by the extracted factors | | | | | | 58.634 |
| Kaiser-Meyer-Olkin Measure of Sampling Adequacy | | | | | | 0.645 |
| Bartlett's Test of Sphericity | | Chi-Square=1977.467; df=78 & P-value<0.001 | | | | |
| Extraction Method | | Principal Component Analysis | | | | |

# CHAPTER FIVE
# FOCUS GROUP DISCUSSION

## 5.1 Issues of Focus Group Discussions (FGD)

As per the plan of the study, six (two in each upazila) Focus Group Discussions (FGDs) on the research topic on" Impact of micro-credit on the livelihoods of clients: A study on Sunamganj District." The FGD were conducted in three upazila during January 2020 to February 2021. The main participants of the FGDs were member of NGOs, lenders (Mohajon) borrowers, non-borrowers and UP Members, data collectors, and Household Heads. The Principal Investigator with the help of Research Assistants conducted the focus group discussions using a well-prepared checklist covering the following issues.

1) Availability of the micro-credit programs in the area;
2) Knowledge about micro-credit programs;
3) Causes of not getting micro-credit from formal sources;
4) Causes of not getting micro-credit from informal sources;
5) Purposes of getting micro-credit programs;
6) Usages of loans;
7) Socio-economic impact of formal micro-credit;
8) Socio-economic impact of informal micro-credit
9) Impact of micro-credit on education and health care expenditure
10) Causes of non-payment of loan in time;
11) Attitudes of borrowers on formal micro-credit programs;
12) Attitudes of borrowers on in-formal micro-credit programs
13) Causes of remaining poor after using micro-credit; and
14) Suggestions for betterment of micro-credit programs.

## 5.2 Outcomes of FGD

**Issue 5.1: Availability of the micro-credit programs in the area**

The formal MCs of NGOs are available but from banks are not easily reachable in the study area. The informal micro-credits from Mohajan, Sammittee are available the informal sources of friends /relatives without interest are available but it depends on persons status and relationship.

**Issue 5.2: Knowledge about micro-credit programs**

In this issue most of the discussants have knowledge about the in-formal micro-credit sources but somebody has less knowledge on formal credit sources and terms of them.

**Issue 5.3: Causes of not getting micro-credit from formal sources**

The participants said that they do not know about some micro-credit programs, nepotism, misappropriation of credits, lack of entry fee/bribe and bureaucratic problem are the main causes of not getting micro-credit from formal sources.

**Issue 5.4: Causes of not getting micro-credit from informal sources**

The discussant told that high rate of interest, social status, non-availability collateral, relationship, lobbing, non-cooperation of local members budget limitation are the main causes of not getting micro-credit from in-formal sources.

**Issue 5.5: Purposes of getting micro-credit programs**

In this issue the participants said that the main purposes of loans are crop cultivation, meeting healthcare expenditure, meeting educational expenses, payment of previous loan, food purchase, construction and repairing of houses mentionable.

**Issue 5.6: Usages of loans**

The borrowers informed that they used loan mostly for consumption of foods, purchasing of agricultural inputs, and repayment of loan, animal purchase and family business.

**Issue 5.7: Socio-economic impact of informal micro-credit**

The participants confirmed that taking informal credits the total income, total expenditure comprising of consumption and investment both has been increased simultaneously with income. The investment in agricultural and productive has been increased highly. The savings has been decreased as they investment increased with income.

**Issue 5.8: Socio-economic impact of formal micro-credit**

Through FGDs, it has been revealed that the agricultural and non-agricultural income and labor sale have been increased with the increase of loans. The consumption of both food and noon food items has been increased while savings increased slightly.

**Issue 5.9: Impact of micro-credit on education and health care expenditure**

The participants of the discussion told that the salient impacts of micro-credit are that both educational and healthcare expenditure has been increased after getting loans surprisingly.

**Issue 5.10: Causes of non-payment of loan in time**

When the participants asked in this regard most of them told that rate of interest, short of installment period, investment loss acute food problem and natural disasters are the main causes of non- payment of formal loans while in case of informal loans the causes are almost same of non-payment of loans. Therefore, there is no differences in the causes of non-payment of formal and informal loans.

**Issue 5.11: Attitudes of borrowers on formal micro-credit programs**

The attitudes towards formal credit are explored by the discussants that the rate of interest is very high, credit period is insufficient and terms were rigid.

**Issue 5.12: Attitudes of borrowers on in- formal micro-credit programs:** Like formal loans the attitudes of the discussant towards informal credit are same *i.e.*, rate of interest is very high, credit period is insufficient, amount of loan is not sufficient as they required.

**Issue 5.13: Causes of remaining poor after using micro-credit**

The participants explored the causes of not come out from vicious cycle of poverty in spite of using loans due to pressure of repayment of loans, loss of investment, not renewal of loans in need, short duration of loans diversion of loans from main purposes, absent of income earning member in the family.

**Issue 5.14: Suggestions for betterment of micro-credit programs**

The most of the participants suggested that for breaking vicious cycle of poverty by micro-credit the duration of loans should be at least five year and the volume of loans must be minimum 500,000 and repayment should at not be less than monthly. The rate of interest should not be more than 5%.

# CHAPTER SIX
## SUMMARY AND CONCLUSION OF THE STUDY

In the previous chapter the results of data analysis have been   presented. Here the summary of the findings is given:

(i) Demographic profile of respondents is that 66.2% respondents of borrowers and 98.7 non-borrowers are head of the family and an average 76.6%. Among the borrowers 32% is husband/wife while 1.3% of non-borrowers and on average 22.2.  In terms of sex 64.7% of borrowers and 92.5% of non-borrowers are male while 35.3% of borrowers and 7.5% of non-borrowers are female.  The age category of respondents showed that 62.5% of borrowers and 66.7% of non –borrowers are 31-50 and on average l 63.8%.  The marital status of the respondents is that 92.2% of borrowers and 92.5% of non –borrowers are married   and on average l 92.3%.  The educational status of the respondents is that   61.9% of borrowers and 61.8% of non–borrowers are up to 5 years schooling    and in total 61.9% following no education 25.4% of borrowers and 18% of non-borrowers and 23% in total. The mean of schooling of the respondents is $3.70 \pm 2.81$ years. The respondents by occupation showed that 24.1%    of borrowers and 25.9% of non –borrowers   are day labor  and in total 24.7% following household working 29.9% borrowers and 4.8% % of non –borrowers   and in total 21.9%,  farming 18.8% of  borrowers and 15.4% % of non –borrowers   and in total 17.7% .In terms of income earner the respondents showed that full time earner is 49.9%   of borrowers and  67.1 % % of non –borrowers   and in total 55.4%  following no work  32.2% of  borrowers and  17.1 % % of non –borrowers   and in total 27.3% , part time worker is 17.9% of borrowers and 15.8% of  non-borrowers and 17.3% in total.  The data showed that among the respondents 2.9% of borrowers and 3.5% of non- borrowers and 3.1% in total are not able.

(ii) The demographic profile of the household members of the respondents showed that the age structure of household members out of 3610 in total showed that 35.5% falls between 0-15 following 28.80% between 16-30, 26% between 31-50 and 9.8% above 51. The marital status of the members above age 16 showed that out of 2329 members 65.7% is married, 28.6% un- married and 5.7% widows and separated. The educational statues of the members above age 7 is that 52.1% has 1-5 years schooling following 19.9% 6-9 years, 15.7 with no education, 10.8% SSC and HSC and only 1.5% graduate and above. The occupation of the members of 16 to 60 aged is that household works consisted of 32.3% following day labor 13.8%, student 12.6%, others 12.6%, service/business 10.8% off-farm activities 10.4% and farming 7.4%. The income earner member above the age of 16 of the households showed that 57.3% members has no work while 30.30% full-time workers, 12.3% part-time workers and others 0.2%. It is observed that 4.5% member is disabling. The composition of household members depicted that male female ratio is 100:93 while female headed is 2.6% and male headed is 97.4% and average family size is 5.6(6) members per household. The dependency ratio is 55% in total where child (0–14) dependency ratio is 48.95% and aged (60+) dependency ratio is 6.05%. In depth analysis of the members of households in terms of education the researcher found the dropout and salient indications on education clarified that 10.4% children aged 6-11 years does not go to primary school while 4.3 children aged 12-18 years does not go to secondary school and 19% adult members aged above 15 never attended school. On the other hand, average schooling of household heads is 3.95 years; husband/wife for 3.32 years and adult members aged 15 and above is 4.96.

(iii) The analysis of housing information of the households of the respondents is explored that 97.1% of borrowers, 93% of non-borrowers and in total 95.8% of both owns housing ownership. The house size is measured in terms of number of room and data showed that 35.1% owns three room following 28.1% two room, 22.2% four room and 14.7% single room while the average number of persons per room is 2.20. There is separate kitchen room in 60.7% houses. There are four types of houses and analysis of data showed that 51.2% holds tin shed roof and wall following 37% tin shed roof and muddy wall, 9.3% tin shed and *Pucca* wall and floor and 2.2% straw roof and bamboo/muddy wall. The findings on sources of cooking fuel showed that 48.4% households use wood/kerosin following 33.3% Straw/Leaf/Husk/Jute stick,16.1% cow dung,2.4% gas and 0.1%others.  The sources of drinking water of the houses is89.8% from tube-well following 9.7 ponds and 0.6% supply water. Electricity coverage in the village or area of the study is 95.5% while the Electricity coverage in the house is 91.3% and the ownership of toilet by the households is 90.3%. Types of toilets used by household members are *Katcha* toilet 57.15 following *pucca* toilet (not water resistant) 35.3% *pucca* toilet (water resistant) 6.3% and open field/others is 1.3%.

(iv) The data on possession of durable assets by households showed that among the top 5 assets 97.1 bed, following 96.8 households owns *Chauk* following, 91.7% mobile phone, 85.7% electric fan, 83.9% table/chair but only 19.5% holds television and 4.1% holds radio.

(v) The information on possession of productive assets by households is classified into two categories one is land and another is o non -land. The land holdings by the households are divided into: (i) homestead land; (ii) cultivable land (own); (iii) cultivable land (leased-in or sharecropped); and (iv) pond. The details of the holdings of land showed that 5.6% holds no homestead land, 79.5% holds 1 to 15 decimals, 7.4% holds 16-50 decimal and 7.4% holds more than 50 decimals. The data on cultivable land (own) explored that 71.1% landless following 15.4% holds more than 50 decimals, 6.7 holds 1 to 15 decimals also 6.7% holds 16-50 decimal. The holding status of cultivable Land (Leased-in or sharecropped) portrayed that 65.5% holds no land following 22.7% more than 50 decimal,7.4%16-50 decimal and 4.3% 1 to 15 decimals. The data on holdings of pond land showed that 96.4% has no pond land while 3.6% holds 1 to 15 Decimal. There are eleven types non-land assets holds by the households as given in table 4.5.2. The data showed that197(41%) borrowers and 72(32%) non borrowers hold livestock following cultivation instruments by 161(33%) borrowers and 59(26%) non borrowers, fishing net 68(14%) borrowers and 43(28%) non borrowers. The combined holdings by both borrowers and non- borrowers of these top three is 268(38%), 220(31%) and 111(16%) respectively.

(vi) The analysis of households' knowledge about micro-credit benefits showed that 95% respondents have knowledge about micro-credit benefit while 85.6% tried to get those benefits. For getting credit firstly 60.2% attempted to nongovernment (MFI/NGO/Insurance) following 13.6% local Money Lender (Mohajan/Private Samittee and 6.9% to more than one sources and 1.5% to government (Banks/Co-operatives). After first attempt the households attempted in second time for getting credit and 9% tried to local Money Lender (Mohajan/Private Samittee following 3.9% Nongovernment (MFI/NGO/Insurance),2.8% Government (Banks/Co-operatives), 0.3% non-interest loan (Relatives/friends/neighbors) and only 0.1% to more than one sources. The respondents seek helped to get credit to different persons or bodies and among these 34.4% went to UP Office following 6.6% relatives/neighbors/friends, 5.5% NGOs, 3.4% government officer, 0.8% UP chairman and 0.7% to UP Member. Among the respondents 2.9% was asked to give money (bribe) for giving micro-credit benefit and 55% told that micro credit remove poverty.

(vii) To find out the factors of exclusion from micro credit program out of fifteen reasons where five causes are significantly mentionable for exclusion from micro credit based on Eigen value (more than 1.00). The Principal Component Analysis (PCA) with Varimax Orthogonal rotation method explained 75.66% of total variation by the extracted factors with 0.703 Kaiser-Meyer-Olkin (KMO) measure of sampling adequacy. It means the sample size is sufficient to explore the factors. The first component comprised of four reasons- (i) no micro-credit in the area; (ii) non availability of collateral; (iii) misappropriation of credit and (iv)business. The second factor is consisted of four reasons- (i) non-cooperation from public delegate of micro-credit institution; (ii) Non-cooperation from local lenders; (iii) Lack of networking or lobbying and (iv) distance from micro-credit from the village.  The third factor included three dimensions- (i) couldn't provide bribe or entry fee, (ii) limitation of budget (according to selector) and non-availability of NID. The fourth factor formed of two dimensions - (i) bureaucratic complexity and (ii) limitation of budget (according to selector). Lastly the fifth factor was made of two dimensions - (i) no political exposure and (ii) nepotism.

(viii) The micro-credit programs available in the study area is investigated   in terms of (i) inclusion month ;(ii) interest rate ;(iii) installment type;(iv) number total installment; (v) duration of loan in month; (vi) collateral or security;(vii) amount of paid loan and interest; and (vii) amount of unpaid loan and interest of both formal and informal loans. The data on month of inclusion depicted that both formal and informal loan are taken during October to December (133+59) following January to March (86+108), July to September (64+32) and April to June (75+14). The insight of interest rate depicted that there is 0% for 20 borrowers of informal credit only. The rate of interest varied between 1% to 25% and more for both of formal and informal credits. The borrowing rate 11% to 15% included maximum formal (143) and informal (160) borrowers following 21% to 25% formal (98) and informal (113), 16% to 20% formal (73) and informal (75), 1% to 10% formal (40) and informal (82) and more than 25% formal (4) and informal (35). The loans are paid in installment basis and the data showed that most of the loans of both formal (287) and informal (9) are on weekly basis following monthly formal (63) and informal (69) and biweekly, quarterly and annually are very small numbers while the number of installments are varied between i to more than 24 and the most of the loans of both formal and informal are in more than 24(formal 290,and informal 12,and total 302) installments following 12(formal 62and informal 43 total 105). It is remarkable that when the lion part of formal loans is paid in more than 24 installments then the informal loans are in single installment. The duration of loans is almost one year case of both formal and informal credits and paid against collateral. The payment of both types of loan is totally successful and both types are loans are paid regularly in time. All the loans are still going and remain a potion is unpaid.

(ix) The borrowers borrow money for different purposes and from pilot survey seventeen (17) purposes of loan are identified and out of them top ten items of purposes in total of both formal and informal loans are purchasing food items is (32%) following crop production (26.2%), payment of previous loan(22.1%), trade/business/industry (20%), rearing cattle/poultry ranked (18.4%), healthcare expenditure (15.7%), repairing cost of houses (13.6%), constructing housing(8.7%) and fish farming/fishing (5.4%). The Principal Component Analysis (PCA) with Varimax Orthogonal rotation method explained 53.22% of total variation by the extracted factors with 0.571 Kaiser-Meyer-Olkin (KMO) measure of sampling adequacy. It means the sample size is sufficient to explore the purposes. The six causes are identified as significant by PCA. The first dimension explained 15.14% variations and composed of purchasing of food items, repairing cost of houses, healthcare expenditure and education. The second dimension explained 9.14% variations with the factors of rearing cattle/poultry, purchasing of livelihood and payment of loan equipment. The third dimension is consisted of crop production and tackling shocks of natural calamities with explaining 8.3% variations. The fourth dimension has explained 7.29% variations composing the tackling shocks of sudden death of HH head and others. The fifth dimension is composed of Sending family member to abroad fish farming/fishing by explaining 7.93% variations. The sixth dimension has explained 6.43% variations having trade/business/industry and daughter/son's marriage purposes. The factors in total explained 53.33% and Kaiser-Meyer-Olkin Measure of Sampling Adequacy is 0.57 which is closed to accept of 0.59.

(x) The respondents were given to mark 24 heads of expenditure and investment out which fourteen heads were recognized. The top seven (7) heads in total of both formal and informal credit are consumption on food (151) following purchasing agricultural inputs (102) payment of loan (93), family enterprises (89), purchasing animals (83) housing improvements (72) and health care (71).

(xi) The impact of informal micro-credit through economic performance based on before-after comparison showed that annual average total income has been increased from Tk. 87393.7 before receiving loan to Tk. 103145.7 after receiving loan than that of before significantly and with this line labor sale has been increased significantly from Tk. 54737.4 to 68075.8 and all the components of total income showed an increase after receiving the loan except business. The annual average total expenditure has been increased from Tk. 90948.8 before receiving loan to Tk. 116220.5 after receiving loan than significantly. The consumption and investment are the main two element of expenditure and both have been increased significantly after credit than that of before. The food and non-food are the two sub-elements of consumption and both have been increased significantly after credit than that of before. There are ten sub-elements of investment and among them agricultural, productive assets and others have been increased significantly. The savings which is the remains after deducting consumption from total income has been decreased insignificantly.

(xii) The information on impact of formal micro-credit through economic performance based on before-after comparison explored that the sources of formal micro-credit are government and non-government banks, co-operatives, microcredit institutions and NGOs. The impact of credit from these sources clarified that total income and its components– agricultural, non–agricultural and labor sale have been increased significantly though income from business and donation have been increased insignificantly. The amount of debt also increased without significance. The total expenditure and its two main components consumption and investment both have been increased significantly. Both the elements of consumption food and non- food has been increased significantly. The elements of investments - education & training, medical, agricultural, family business, household development and others have been increased significantly but savings increased insignificantly.

(xiii) The impact of micro-credit in terms of formal and informal credit receiving households based on DID method showed the over-all impact of both formal and informal credit. It portrayed that total income; total expenditure and investment have been increased 13.57%, 10.39% and 26.17%. All the elements of total income have been increased except debt which has been decreased by 2.39%. But the decrease in debt is the good sign of positive impact of debt. Consumption of food has been increased but non-food has been decreased. All the elements of investment have been increased except some factors. The savings has been decreased due excess increase in investment.

(xiv) Percentage change of households' benefits using micro-credit based on DID method showed that total income has been increased by 3.25% over the study period. All the elements of total income have been increased except labor sale while it has been decreased. It means that some of the borrowers have become self-reliance by using debt as debt has been also increased. The total expenditure has been increased by 8.76% over the study period. The consumption expenditure is one of the elements of total expenditure has been decreased by 4.42% while investment (capital) expenditure has been increased at show high rate (31.3%) as a result saving has been decreased by 11.8%.

(xv) The percentage distribution of households' food insecurity during the study period showed that food security has been classified in terms of number of meals taken the respondents in a day and based on this criterion we divided the respondents into – (i) some periods of hunger (normal food insecurity) (iii) two meals a day throughout year (no food insecurity) and (iii) three meals a day throughout year (no food insecurity). The data showed that normal food insecurity of both borrowers and non- borrowers have been decreased during the study period and in total it has been decreased from 22.1% to 13% similarly no food insecurity of both borrowers and non- borrowers have been decreased during the study period and in total it has been decreased from 28.1% to 24.7% but finally no food insecurity has been increased of both borrowers and non- borrowers over the study period and in total it has been increased from 49.8% to 62.3%. Therefore, the use of micro credit ensures food security and reduces food insecurity.

(xvi) Percentage distribution of households self-assessed socio-economic status over the study period showed that the role of micro -credit graduation of households from lower class to upper class. For this purpose, the households have been divided into- (i) extremely poor; (ii) moderately poor ;(iii) poor; (iv) middle class; and (v) rich. The analysis of data found that both borrowers and non-borrowers under extremely poor has been decreased over the study period and in total it has been decreased from 25.5% to 13.9% similarly both borrowers and non-borrowers under moderately poor has been decreased over the study period and in total it has been decreased from 31.7% to 26.6% while both borrowers and non-borrowers under poor has been increased over the study period and in total it has been increased from 36.6% to 50.4% and both borrowers and non-borrowers under middle class has been increased over the study period and in total it has been increased from 4.9% to 7.9% and yet there is no change in rich category. So, microcredit helps to come out from extremely poor to moderately poor and from to moderately poor to poor and from moderately poor to poor be middle class and in long run it will help to be rich from middle class.

(xvii) The impact of micro-credit on changing educational and healthcare expenditure in 2019 /20compared to 2016/17 showed that education expenditure has been increased in case of both borrowers and non –borrowers over the study period and in total it has been increased from 18.4% to 58.8% while no change has been decreased of both borrowers and non –borrowers and in total it has been decreased from 55.5% to 38%. in the same time of both borrowers and non –borrowers over the study period have been decreased which in total from 26.1% to 3.2%. The data showed that health expenditure has been increased in case of both borrowers and non –borrowers over the study period and in total it has been increased from 29% to 57.6% while no change has been decreased of both borrowers and non –borrowers and in total it has been decreased from 49.2% to 33.5% in the same time of both borrowers and non –borrowers over the study period has been decreased which in total from 21.7% to 8.8%. Therefore, micro-credit has positive impact on changing on both education and health development.

(xix) In order to identify the factors responsible for non-payment of loan we have listed 15 factors measured in term of percentage of the respondents and the analysis of data showed that that among the 15 factors rate of interest is very high is agreed by 64.2% of formal borrowers following 60.1% of installment period is very short, 50.8% of investment loss 47.8% of acute food problem and natural calamities 46.9%. While in case of formal credit 84.3% of rate of interest is very high following 77.2 of misappropriation of loan, 75.6% of Installment period is very short, 72.4% of both acute food problem and investment loss and 63.8% of natural calamities. In total, rate of interest is very high stood (69.5%), installment period is very short is in second (64.1%) investment loss is in third (56.5%), acute food problem is in forth (54.2%), and medical treatment/medicine is in the fifth (49.5). Therefore, in case of both formal and in-formal cases the top most five factors of non-payment of loan are about same and they are rate of interest is very high, installment period is very short, investment, acute food problem, and natural calamities.

Through the PCA with Varimax Orthogonal rotation method explained 66.48 % of total variation by the extracted factors with 0.791 Kaiser-Meyer-Olkin (KMO) measure of sampling adequacy. It means the sample size is sufficient to explore the factors. The first factor is composed of six factors with 32.14(around 50% of 66.48%) percentage of total variation explained and they are (i) acute food problem; (ii) insufficient of loan for investment;(iii) duration of loan is short for return from investment; (iv) installment period is very short; (v) rate of interest is very high and (vi) renewal of loan is unavailable. The second factor is composed of three factors with 16.19(around 25% of 66.48%) percentage of total variation explained and they are (i) expenditure for marriage of son/daughter; (ii) family legal problems and expenditure and (iii) unexpected accident. The third factor is composed of three factors with 10.61 (around 15% of 66.48%) percentage of total variation explained (i) medical treatment/medicine; (ii) natural calamities and (iii) crop's failure lastly the fourth factor composed of two factors with 7.55 (more than 10% of 66.48%) percentage of total variation explained and they are investment loss and misappropriation of loan.

(xx) To measure the attitude of respondents towards micro-credit we have listed 16 factors measured in term of percentage of the respondents with respect to disagree, neutral and agree. The analysis of data showed that among the 16 factors in case of formal credit among the top five factors showed that the rate of interest of micro-credit is reasonable is disagreed by 74.6% and agreed only by 17% following duration of credit is sufficient disagreed by 61.7% and agreed only by 21.2%, amount of credit is sufficient disagreed by 47.5% and agreed only by 35.5 %, by micro-finance your savings has increased is disagreed by 37.4% and agreed only by 32.4% and terms and conditions are not rigid disagreed by 30.4% and agreed only by 57.5%. Side by side in case of informal credit portrayed that among the 16 factors the top five factors showed that the rate of interest of micro-credit is reasonable is disagreed by 94.5% and agreed only by 3.1% following duration of credit is sufficient disagreed by 78% and agreed only by 18.9%, amount of credit is sufficient disagreed by 77.2% and agreed only by 19.7%, by micro-finance your savings has increased is disagreed by 66.9% and agreed only by 9.4 % and terms and conditions are not rigid disagreed by 57.5% and agreed only by 37.5%. Over-all we found that the top five factors are the rate of interest of micro-credit is reasonable is disagreed by 79.8% and agreed only by 13.4% following duration of credit is sufficient disagreed by 66.0% and agreed only by 20.6%, amount of credit is sufficient disagreed by 55.3% and agreed only by 31.3%, by micro-finance your savings has increased is disagreed by 45.2% and agreed only by 26.4 % and terms and conditions are not rigid disagreed by 37.5% and agreed only by 30.9%. Therefore, there are very many similarities of attitude towards micro-credit of the borrowers in terms of formal credit, informal credit and in over-all of the two. The PCA analysis showed that through the Varimax rotated factor loading four factors are identified with 60.885% total variation explained after factors are extracted. The first factor has explained 26.316% of variations with six factors and they are (i) amount of credit is sufficient; (ii) by micro-finance your income has increased; (iii) by micro-finance your savings has increased; (iv) micro-finance is helping you in better access to education; (v) micro-finance is helping you in better access to healthcare; and (vi) due to micro-finance, employment opportunities have been increased. The second factor with explaining 14.154% of variation is composed of four sub factors which are (i) local loans are easier to get than MFIs; (ii) local lenders are friendly than MFIs; (iii) cost of local loans is lower than MFIs; and (iv) terms and conditions of local loans are easier than MFIs. The third principal factor with explaining 12.185% of variation is composed of three factors which are (i) rate of interest of micro-credit is reasonable; (ii) duration of credit is sufficient; and (iii) by micro-finance your food security has increased. The fourth principal factor with explaining 8.230% of variation is composed of three factors which are (i) terms and conditions are not rigid; (ii) micro-finance is helping you in better financial situation of your; and (iii) operational assistance received from MFIs was helpful to run the business.

(xxi) The analysis of causes of not over-coming from vicious cycle of poverty of not overcoming from vicious cycle of poverty showed that thirteen causes are listed and the respondents marked on three-point scales which are agree, neutral and disagree. It is found that the most five causes agreed by non-borrowers are natural calamity (89%), following not getting loan in time (77.2) loss of investment (75.9), high interest (75.4%) and number of dependent members are high (69.7%). In the same way the most five causes agreed by borrowers are high interest (89.1%) following pressure of loan payment (76.5%), duration of loan is insufficient (71.5%), Insufficient loan (59%), natural calamity (58.8%) and loss of investment (58.1%). In total the top five agreed causes of not over-coming from vicious cycle of poverty are high interest (84.7%) following pressure of loan payment (71.2%), natural calamity (68.4%), duration of loan is insufficient (64.7%) and loss of investment (63.8%).

**Conclusions and Policy Implications:** The study through first hand data regarding the impact of micro-credit on the livelihoods of the clients *i.e.,* borrowers have been analyzed through descriptive as well as inferential statistics and found that microcredit of all sources has positive impact on the livelihoods of the clients in terms of socioeconomic factors viz. expenditure, consumption, education, healthcare, food security, holdings of both durable and productive assets. On the basis of quantitative data analysis and qualitative (FGD) analysis for productive outcome from micro-credits the study recommended the followings:

(i)     In order to break the vicious cycle of poverty of the clients by using micro-credit the duration of loans should be at least five year and the volume of loans must be minimum 500,000 and repayment should at not be less than monthly. The rate of interest should not be more than 5%.

(ii)    The clients should be identified indifferently for formal loans and the informal lenders should be listed under the program to reduce their despotism.

(iii)   Further research on lenders profitability may be done.

*Annex-1: Appendix Tables*

**Appendix table 4.1: Demographic Characteristics of Respondents' by Borrowing Status.**

| Characteristics | Type of Household | | Z-statistic | P-value | Both |
|---|---|---|---|---|---|
| | Borrower | Non-borrower | | | |
| **Relation with HH Head** | | | | | |
| Head | 66.2 | 98.7 | -9.559 | <0.001 | 76.6 |
| Husband/Wife | 32.0 | 1.3 | 9.202 | <0.001 | 22.2 |
| Son/Daughter | 0.6 | 0.0 | 1.172 | 0.201 | 0.4 |
| Father/Mother | 0.6 | 0.0 | 1.172 | 0.201 | 0.4 |
| Others | 0.6 | 0.0 | 1.172 | 0.201 | 0.4 |
| **Sex of Respondent** | | | | | |
| Male | 64.7 | 92.5 | -7.853 | <0.001 | 73.6 |
| Female | 35.3 | 7.5 | 7.853 | <0.001 | 26.4 |
| **Age of Respondent** | | | | | |
| 16-30 | 19.8 | 10.1 | 3.239 | 0.002 | 16.7 |
| 31-50 | 62.5 | 66.7 | -1.089 | 0.221 | 63.8 |
| 51-60 | 12.0 | 18.0 | -2.159 | 0.039 | 13.9 |
| Above 60 | 5.8 | 5.3 | 0.270 | 0.385 | 5.6 |
| Mean age | $41.97 \pm 11.15$ | $44.04 \pm 10.41$ | -2.420 | 0.021 | $42.63 \pm 10.96$ |
| **Marital Status** | | | | | |
| Married | 92.2 | 92.5 | -0.140 | 0.395 | 92.3 |
| Unmarried | 3.7 | 0.9 | 2.112 | 0.043 | 2.8 |
| Widow | 4.1 | 6.1 | -1.172 | 0.201 | 4.8 |
| Separated/Divorced | 0.0 | 0.4 | -1.394 | 0.151 | 0.1 |
| **Educational Status** | | | | | |
| No education | 25.4 | 18.0 | 2.189 | 0.036 | 23.0 |
| 1-5 years of schooling | 61.9 | 61.8 | 0.026 | 0.399 | 61.9 |
| 6-9 years of schooling | 9.7 | 17.1 | -2.829 | 0.007 | 12.1 |
| SSC / HSC | 2.9 | 2.6 | 0.226 | 0.389 | 2.8 |
| Graduate and above | 0.2 | 0.4 | -0.485 | 0.355 | 0.3 |
| Mean years of schooling | $3.45 \pm 2.77$ | $4.24 \pm 2.82$ | -3.509 | 0.001 | $3.70 \pm 2.81$ |
| **Occupation** | | | | | |
| Farming | 18.8 | 15.4 | 1.109 | 0.216 | 17.7 |
| Day laborer | 24.1 | 25.9 | -0.520 | 0.349 | 24.7 |
| Off-farm activities | 8.9 | 25.9 | -6.041 | <0.001 | 14.3 |
| Service/Business | 12.8 | 19.7 | -2.406 | 0.022 | 15.0 |
| Student | 0.4 | 0.0 | 0.956 | 0.253 | 0.3 |
| Household Work | 29.9 | 4.8 | 7.562 | <0.001 | 21.9 |
| Others | 5.2 | 8.3 | -1.602 | 0.111 | 6.2 |
| **Income Earner** | | | | | |
| Full time | 49.9 | 67.1 | -4.309 | <0.001 | 55.4 |
| Part time | 17.9 | 15.8 | 0.693 | 0.314 | 17.3 |
| No work | 32.2 | 17.1 | 4.218 | <0.001 | 27.3 |
| **Disability Status** | | | | | |
| Yes | 2.9 | 3.5 | -0.432 | 0.363 | 3.1 |
| No | 97.1 | 96.5 | 0.432 | 0.363 | 96.9 |
| **Total(n)** | **485** | **228** | | | **713** |

**Appendix table 4.2: Demographic Profile of Household (HH) Members of Respondents.**

| Characteristics | Sex of HHs members | | Both (%) |
|---|---|---|---|
| | Male (%) | Female (%) | |
| **Classification by Age** | | | |
| 0-15 | 35.6 | 35.4 | 35.5 |
| 16-30 | 27.5 | 30.2 | 28.8 |
| 31-50 | 25.5 | 26.5 | 26.0 |
| 51-60 | 6.1 | 4.4 | 5.3 |
| Above 60 | 5.3 | 3.5 | 4.5 |
| Mean age | 27.32 ± 18.44 | 25.90 ± 17.02 | 26.64 ± 18.00 |
| Total (n) | 1880 | 1730 | 3610 |
| **Marital Status (Age 16 years or more)** | | | |
| Married | 63.0 | 68.6 | 65.7 |
| Unmarried | 34.2 | 22.6 | 28.6 |
| Widow | 2.6 | 8.4 | 5.4 |
| Separated/Divorced | 0.2 | 0.4 | 0.3 |
| Total (n) | 1211 | 1118 | 2329 |
| **Educational Status (Age 7 years or more)** | | | |
| No education | 13.4 | 18.2 | 15.7 |
| 1-5 years of schooling | 53.5 | 50.6 | 52.1 |
| 6-9 years of schooling | 21.4 | 18.4 | 19.9 |
| SSC / HSC | 10.2 | 11.3 | 10.8 |
| Graduate and above | 1.5 | 1.4 | 1.5 |
| Total (n) | 1690 | 1569 | 3259 |
| **Occupation (for age 16-60 years)** | | | |
| Farming | 15.1 | 0.1 | 7.8 |
| Day laborer | 26.2 | 0.8 | 13.8 |
| Off-farm activities | 19.6 | 0.8 | 10.4 |
| Service/Business | 17.3 | 4.1 | 10.8 |
| Student | 11.2 | 14.1 | 12.6 |
| Household Work | 0.0 | 66.3 | 32.3 |
| Others | 10.6 | 13.9 | 12.2 |
| Total (n) | 1111 | 1057 | 2168 |
| **Income Earner (Age 16 years or more)** | | | |
| Full time | 51.5 | 7.2 | 30.3 |
| Part time | 18.4 | 5.6 | 12.3 |
| No work | 30.1 | 86.8 | 57.3 |
| Otherwise | 0.0 | 0.4 | 0.2 |
| Total (n) | 1211 | 1118 | 2329 |
| **Disability Status of Household member** | | | |
| Yes | 4.5 | 4.5 | 4.5 |
| No | 95.5 | 95.5 | 95.5 |
| Total(n) | 1880 | 1730 | 3610 |
| **Household Composition** | | | |
| Sex Ratio | 93 women per 100 men | | |
| Female-headed household (percent) | 2.6 | | |
| **Dependency Ratio (%)** | | | |
| Child (0–14) dependency ratio (divided by the labor force) | 48.95 | | |
| Aged (60+) dependency ratio (divided by the labor force) | 6.05 | | |
| Total dependency ratio (child + aged dependency ratio) | 55.00 | | |
| **Average Family size** | 5.06 | | |

**Appendix table 4.3: Housing Condition and Sanitation Facilities by Micro-credit Status.**

| Indicators | Type of the Households | | | | Total | |
|---|---|---|---|---|---|---|
| | Borrower | | Non-borrower | | | |
| | Number | % | Number | % | Number | % |
| **Ownership of Living House** | | | | | | |
| Yes | 471 | 97.1 | 212 | 93.0 | 683 | 95.8 |
| No | 14 | 2.9 | 16 | 7.0 | 30 | 4.2 |
| **Number of Rooms** | | | | | | |
| One rooms | 58 | 12.0 | 47 | 20.6 | 105 | 14.7 |
| Two rooms | 145 | 29.9 | 55 | 24.1 | 200 | 28.1 |
| Three rooms | 167 | 34.4 | 83 | 36.4 | 250 | 35.1 |
| Four or more rooms | 115 | 23.7 | 43 | 18.9 | 158 | 22.2 |
| **Average person per room** | 2.20 | | 2.20 | | 2.20 | |
| **Separate Kitchen in the Household** | | | | | | |
| Yes | 313 | 64.5 | 120 | 52.6 | 433 | 60.7 |
| No | 172 | 35.5 | 108 | 47.4 | 280 | 39.3 |
| **Regular income in family** | | | | | | |
| Yes | 242 | 49.9 | 112 | 49.1 | 354 | 49.6 |
| No | 243 | 50.1 | 116 | 50.9 | 359 | 50.4 |
| **Type of main house** | | | | | | |
| Straw roof and bamboo/muddy wall | 13 | 2.7 | 3 | 1.3 | 16 | 2.2 |
| Tin shed roof and muddy wall | 161 | 33.2 | 103 | 45.2 | 264 | 37.0 |
| Tin shed roof and wall | 285 | 58.8 | 80 | 35.1 | 365 | 51.2 |
| Building (Tin shed and *Pucca*) | 26 | 5.4 | 40 | 17.5 | 66 | 9.3 |
| **Source of cooking fuel** | | | | | | |
| Wood/Kerosene | 247 | 50.9 | 98 | 43.0 | 345 | 48.4 |
| Gas | 5 | 1.0 | 12 | 5.3 | 17 | 2.4 |
| Straw/Leaf/Husk/Jute stick | 172 | 35.5 | 63 | 27.6 | 235 | 33.0 |
| Cow dung | 61 | 12.6 | 54 | 23.7 | 115 | 16.1 |
| Others | 0 | 0.0 | 1 | 0.4 | 1 | 0.1 |
| **Main source of drinking water** | | | | | | |
| Supply water | 4 | 0.8 | 0 | 0.0 | 4 | 0.6 |
| Tube-well | 432 | 89.1 | 208 | 91.2 | 640 | 89.8 |
| Pond/Well | 49 | 10.1 | 20 | 8.8 | 69 | 9.7 |
| **Electricity coverage in the village or area** | | | | | | |
| Yes | 462 | 95.3 | 219 | 96.1 | 681 | 95.5 |
| No | 23 | 4.7 | 9 | 3.9 | 32 | 4.5 |
| **Electricity coverage in the house** | | | | | | |
| Yes | 438 | 90.3 | 213 | 93.4 | 651 | 91.3 |
| No | 47 | 9.7 | 15 | 6.6 | 62 | 8.7 |
| **Ownership of Toilet** | | | | | | |
| Yes | 447 | 92.2 | 197 | 86.4 | 644 | 90.3 |
| No | 38 | 7.8 | 31 | 13.6 | 69 | 9.7 |
| **Type of Toilet used by HH members** | | | | | | |
| *Pucca* toilet (water resistant) | 13 | 2.7 | 32 | 14.0 | 45 | 6.3 |
| *Pucca* toilet (not water resistant) | 185 | 38.1 | 67 | 29.4 | 252 | 35.3 |
| *Katcha* toilet | 278 | 57.3 | 129 | 56.6 | 407 | 57.1 |
| Open field/Others | 9 | 1.9 | 0 | 0.0 | 9 | 1.3 |
| **Total (n)** | **485** | | **228** | | **713** | |

**Appendix table 4.5.1: Landholdings by Study Households.**

| Type of Land | Type of the Households | | | | Total | |
|---|---|---|---|---|---|---|
| | Borrower | | Non-borrower | | | |
| | No. | % | No. | % | No. | % |
| **Homestead Land** | | | | | | |
| No Homestead Land | 23 | 4.7 | 17 | 7.5 | 40 | 5.6 |
| 1 to 15 Decimal | 389 | 80.2 | 178 | 78.1 | 567 | 79.5 |
| 16-50 Decimal | 37 | 7.6 | 16 | 7.0 | 53 | 7.4 |
| More than 50 Decimal | 36 | 7.4 | 17 | 7.5 | 53 | 7.4 |
| Average ± SD | 15.34 ± 32.43 | | 13.52 ± 28.77 | | 14.77 ± 31.32 | |
| **Cultivable Land (Own)** | | | | | | |
| Landless | 343 | 70.7 | 164 | 71.9 | 507 | 71.1 |
| 1 to 15 Decimal | 33 | 6.8 | 15 | 6.6 | 48 | 6.7 |
| 16-50 Decimal | 37 | 7.6 | 11 | 4.8 | 48 | 6.7 |
| More than 50 Decimal | 72 | 14.8 | 38 | 16.7 | 110 | 15.4 |
| Average ± SD | 66.07 ± 63.65 | | 103.56 ± 159.32 | | 77.72 ± 104.35 | |
| Median | 57.50 | | 60.00 | | 60.00 | |
| **Cultivable Land (Leased-in or sharecropped)** | | | | | | |
| No Sharecropped land | 302 | 62.3 | 165 | 72.4 | 467 | 65.5 |
| 1 to 15 Decimal | 21 | 4.3 | 10 | 4.4 | 31 | 4.3 |
| 16-50 Decimal | 45 | 9.3 | 8 | 3.5 | 53 | 7.4 |
| More than 50 Decimal | 117 | 24.1 | 45 | 19.7 | 162 | 22.7 |
| Average ± SD | 107.20 ± 92.77 | | 116.92 ± 97.52 | | 109.69 ± 93.91 | |
| **Pond land** | | | | | | |
| Don't have | 466 | 96.1 | 221 | 96.9 | 687 | 96.4 |
| 1 to 15 Decimal | 17 | 3.5 | 5 | 2.2 | 22 | 3.6 |
| Average ± SD | 6.26 ± 7.24 | | 11.14 ± 10.85 | | 7.58 ± 8.42 | |
| **Total (n)** | **485** | | **228** | | **713** | |

**Appendix table 4.6: Household's Knowledge about Micro-credit Benefits.**

| Indicator | Number | Percentage |
|---|---|---|
| **Knowledge about Micro-credit benefit** | | |
|   Yes | 683 | 95.8 |
|   No | 30 | 4.2 |
| **Trying about getting Micro-credit benefit** | | |
|   Yes | 610 | 85.6 |
|   No | 103 | 14.4 |
| **First attempt for getting Micro-credit benefit** | | |
|   Government (Banks/Co-operatives) | 11 | 1.5 |
|   Nongovernment (MFI/NGO/Insurance) | 429 | 60.2 |
|   Local Money Lender (Mohajan/Private Samittee) | 97 | 13.6 |
|   Non-interest Loan (Relatives/friends/neighbors) | 24 | 3.4 |
|   More than one sources | 49 | 6.9 |
| **Second attempt for getting Micro-credit benefit** | | |
|   Government (Banks/Co-operatives) | 20 | 2.8 |
|   Nongovernment (MFI/NGO/Insurance) | 28 | 3.9 |
|   Local Money Lender (Mohajan/Private Samittee) | 64 | 9.0 |
|   Non-interest Loan (Relatives/friends/neighbors) | 2 | 0.3 |
|   More than one sources | 1 | 0.1 |
| **Ask for help to get Micro-credit benefit** | | |
|   UP Chairman | 6 | 0.8 |
|   UP Member | 5 | 0.7 |
|   UP Office | 245 | 34.4 |
|   Government Officer | 24 | 3.4 |
|   Relatives/Neighbors/Friends | 47 | 6.6 |
|   NGO | 39 | 5.5 |
| **Ask for money for giving Micro-credit benefit** | | |
|   Yes | 21 | 2.9 |
|   No | 357 | 50.1 |
| **Micro-credits help to remove poverty** | | |
|   Yes | 392 | 55.0 |
|   No | 32 | 4.5 |
| **Total (n)** | **713** | |

**Appendix table 4.7: Causes of for Excluded from Micro-credit Program by Non-borrower Participation.**

| Name of the reasons | Non-borrower Households | | | | | |
|---|---|---|---|---|---|---|
| | **Household engaged** | Strongly disagree | Disagree | No Comment | Agree | Strongly agree |
| Bureaucratic complexity | 210 | 0.3 | 7.3 | 13.6 | 3.2 | 5.0 |
| Limitation of budget (according to selector) | 209 | 5.2 | 6.0 | 11.8 | 2.5 | 3.8 |
| Couldn't provide bribe or entry fee | 209 | 5.3 | 8.6 | 10.1 | 3.9 | 1.4 |
| No political exposure | 209 | 5.6 | 4.3 | 9.4 | 6.7 | 3.2 |
| Didn't have any idea about such program | 210 | 9.3 | 10.7 | 3.8 | 3.5 | 2.2 |
| Nepotism | 200 | 4.1 | 3.8 | 14.0 | 4.3 | 1.8 |
| Non-cooperation from public delegate of Micro-credit institution | 210 | 0.8 | 10.9 | 4.8 | 11.1 | 1.8 |
| Non-cooperation from local lenders | 209 | 0.7 | 5.6 | 8.1 | 13.0 | 1.8 |
| Non-availability of NID | 209 | 10.8 | 9.8 | 7.2 | 1.1 | 0.4 |
| Lack of networking or lobbying | 208 | 3.1 | 4.3 | 8.6 | 11.4 | 1.8 |
| Distance from Micro-credit from the village | 207 | 2.4 | 14.0 | 4.5 | 7.3 | 0.8 |
| No Micro-credit in the area | 154 | 7.9 | 11.2 | 1.5 | 0.1 | 0.8 |
| Non availability of collateral | 152 | 0.6 | 5.3 | 3.8 | 9.4 | 2.2 |
| Misappropriation of credit | 153 | 1.5 | 2.1 | 5.0 | 11.2 | 1.5 |
| Biasness | 130 | 0.1 | 6.2 | 10.9 | 0.7 | 0.3 |
| Others | 76 | 0.6 | 0.1 | 6.3 | 1.5 | 0.1 |

**Appendix table 4.8: Profile of Micro-credit Programs.**

| Profile of Micro-credits | Type of Micro-credit | | | | Total | |
|---|---|---|---|---|---|---|
| | Formal | | Informal | | | |
| | N | Average | N | Average | N | Average |
| **Inclusion Months** | | | | | | |
| January to March | 86 | 21244.19 | 22 | 33818.18 | 108 | 23805.56 |
| April to June | 75 | 23600.00 | 14 | 38571.43 | 89 | 25955.06 |
| July to September | 64 | 34390.62 | 32 | 40718.75 | 96 | 36500.00 |
| October to December | 133 | 24112.78 | 59 | 18847.46 | 192 | 22494.79 |
| **Interest Rate** | | | | | | |
| No interest (0%) | - | - | 20 | 13750.00 | 20 | 13750.00 |
| 1% to 10% | 40 | 34600.00 | 42 | 27642.86 | 82 | 31036.59 |
| 11% to 15% | 143 | 21041.96 | 17 | 13882.35 | 160 | 20281.25 |
| 16% to 20% | 73 | 24041.10 | 2 | 17500.00 | 75 | 23866.67 |
| 21% to 25% | 98 | 28132.65 | 15 | 36133.33 | 113 | 29194.69 |
| More than 25% | 4 | 25000.00 | 31 | 46774.19 | 35 | 44285.71 |
| **Installment Type** | | | | | | |
| Weekly | 287 | 23644.60 | 9 | 24111.11 | 296 | 23658.78 |
| Biweekly | 3 | 20000.00 | - | - | 3 | 20000.00 |
| Monthly | 63 | 32365.08 | 69 | 32449.28 | 132 | 32409.09 |
| Quarterly | 2 | 25000.00 | 14 | 15928.57 | 16 | 17062.50 |
| Annually | 3 | 23333.33 | 35 | 29142.86 | 38 | 28684.21 |
| **Total Installment** | | | | | | |
| One time | 3 | 23333.33 | 58 | 22603.45 | 61 | 22639.34 |
| Up to 12 times | 62 | 32241.94 | 43 | 28906.98 | 105 | 30876.19 |
| From 13 to 24 times | 3 | 25000.00 | 14 | 49857.14 | 17 | 45470.59 |
| More than 24 times | 290 | 23658.62 | 12 | 37250.00 | 302 | 24198.68 |
| **Duration of Loan** | | | | | | |
| Six months | 1 | 5000.00 | 3 | 36666.67 | 4 | 28750.00 |
| One year | 356 | 25140.45 | 110 | 26281.82 | 466 | 25409.87 |
| Two years | 1 | 50000.00 | 14 | 49857.14 | 15 | 49866.67 |
| **Collateral Type** | | | | | | |
| Collateral | 353 | 25178.47 | 112 | 24973.21 | 465 | 25129.03 |
| Non collateral | 5 | 23400.00 | 15 | 60133.33 | 20 | 50950.00 |
| **Paid Loan** | | | | | | |
| Total paid | 358 | 14401.04 | 127 | 9476.50 | 485 | 13111.52 |
| Principal | 358 | 12444.15 | 127 | 7974.95 | 485 | 11273.86 |
| Interest | 358 | 1956.90 | 127 | 1501.54 | 485 | 1837.66 |
| **Unpaid Loan** | | | | | | |
| Total unpaid loan | 358 | 15232.06 | 127 | 26972.63 | 485 | 18306.39 |
| Unpaid principal | 358 | 12709.49 | 127 | 21151.03 | 485 | 14919.95 |
| Unpaid interest | 358 | 2522.57 | 127 | 5821.60 | 485 | 3386.44 |
| **Total** | **358** | **25153.63** | **127** | **29125.98** | **485** | **26193.81** |

**Appendix table 4.9.1: Purpose of Loan of the Study Households.**

| Purpose of Loan | Micro-credit Types | | | | Total | |
|---|---|---|---|---|---|---|
| | **Formal** | | **Informal** | | | |
| | **Yes (%)** | **No (%)** | **Yes (%)** | **No (%)** | **Yes (%)** | **No (%)** |
| Purchasing of food items | 31.8 | 68.2 | 32.3 | 67.7 | 32.0 | 68.0 |
| Crop production | 22.1 | 77.9 | 37.8 | 62.2 | 26.2 | 73.8 |
| Rearing cattle/poultry | 22.1 | 77.9 | 7.9 | 92.1 | 18.4 | 81.6 |
| Sending family member to abroad | 2.0 | 98.0 | 0.8 | 99.2 | 1.6 | 98.4 |
| Trade/Business/Industry | 21.8 | 78.2 | 15.0 | 85.0 | 20.0 | 80.0 |
| Fish farming/Fishing | 4.7 | 95.3 | 7.1 | 92.9 | 5.4 | 94.6 |
| Daughter/son's marriage | 1.7 | 98.3 | 2.4 | 97.6 | 1.9 | 98.1 |
| Constructing housing | 9.5 | 90.5 | 6.3 | 93.7 | 8.7 | 91.3 |
| Tackling shocks of natural calamities | 2.2 | 97.8 | 1.6 | 98.4 | 2.1 | 97.9 |
| Tackling shocks of sudden death of HH head | 0.3 | 99.7 | - | 100.0 | 0.2 | 99.8 |
| Purchasing of livelihood equipment | 8.9 | 91.1 | 8.7 | 91.3 | 8.9 | 91.1 |
| Payment of loan | 24.9 | 75.1 | 14.2 | 85.8 | 22.1 | 77.9 |
| Repairing cost of houses | 15.9 | 84.1 | 7.1 | 92.9 | 13.6 | 86.4 |
| Healthcare expenditure | 16.2 | 83.8 | 14.2 | 85.8 | 15.7 | 84.3 |
| Education | 8.7 | 91.3 | 4.7 | 95.3 | 7.6 | 92.4 |
| Others | 17.0 | 83.0 | 12.6 | 87.4 | 15.9 | 84.1 |
| **Total (n)** | **358** | | **127** | | **485** | |

**Appendix table 4.11.1: Impact of Informal Micro-credit through Economic Performance   based on before-after Comparison**

| Sources of Economic Performance | Three year (in 2016/17) before survey | | | Survey point (in 2019/20) | | | t-test |
|---|---|---|---|---|---|---|---|
| | HHs | Average | SD | HHs | Average | SD | |
| **Total Income** | **127** | **87393.7** | **44698.4** | **127** | **103145.7** | **54935.3** | -2.51** |
| Agricultural | 84 | 40267.9 | 20091.7 | 90 | 39855.6 | 20308.6 | 0.13 |
| Non-agricultural | 37 | 38189.2 | 30380.8 | 40 | 42625.0 | 31954.0 | -0.62 |
| Labor Sale | 99 | 54737.4 | 27218.7 | 99 | 68075.8 | 42234.2 | -2.64*** |
| Business | 20 | 43450.0 | 29324.8 | 24 | 42416.7 | 21779.4 | 0.13 |
| Donation/Begging | 06 | 2583.3 | 1744.0 | 07 | 7142.9 | 10318.7 | -1.06 |
| Debt | 78 | 17397.4 | 14015.1 | 114 | 21263.2 | 23302.6 | -1.31 |
| **Total Expenditure** | **127** | **90948.8** | **40386.8** | **127** | **116220.5** | **53284.5** | -4.26*** |
| **Consumption** | 127 | **56252.0** | 17859.2 | 127 | **66952.8** | 23963.2 | -4.04*** |
| Food | 127 | 45842.5 | 14116.0 | 127 | 53409.5 | 17832.7 | -3.75*** |
| Non-Food | 127 | 10409.5 | 10409.5 | 127 | 13543.3 | 8158.9 | -2.67*** |
| **Investment** | 125 | **35252.0** | 28981.5 | 127 | **49267.7** | 35744.7 | -3.42*** |
| Education & Training | 80 | 10068.8 | 7505.2 | 91 | 11351.6 | 8719.5 | -1.02 |
| Medical | 124 | 9032.3 | 10533.7 | 124 | 10975.8 | 9109.5 | -1.55 |
| Agricultural | 99 | 14015.2 | 10486.5 | 101 | 17108.9 | 13480.1 | -1.81* |
| Family Business | 09 | 13444.4 | 8125.7 | 14 | 19642.9 | 14008.8 | -1.20 |
| HH Development | 49 | 10051.0 | 12169.1 | 69 | 12681.2 | 10117.7 | -1.28 |
| Productive Asset | 12 | 8583.3 | 3941.8 | 15 | 11733.3 | 5035.1 | -1.77* |
| Durable Goods | 10 | 9600.0 | 4141.9 | 13 | 10000.0 | 4899.0 | -0.21 |
| House Repairing | 36 | 7458.3 | 12528.5 | 50 | 11320.0 | 9685.8 | -1.61 |
| Land Purchasing | 02 | 12500.0 | 14849.2 | 01 | 3000.0 | 0.0 | 0.52 |
| Other Investment | 57 | 8421.1 | 8220.1 | 65 | 15153.9 | 16855.5 | -2.74*** |
| Savings | 08 | 3750.0 | 6267.8 | 06 | 3166.7 | 2041.2 | 0.22 |
| **Total Households** | **127** | | | **127** | | | |

**Appendix table 4.11 .2: Impact of Formal Micro-credit through Economic Performance based on before-after Comparison.**

| Sources of Economic Performance | Three year (in 2016/17) before survey | | | Survey point (in 2019/20) | | | t-test |
|---|---|---|---|---|---|---|---|
| | HHs | Average | SD | HHs | Average | SD | |
| **Total Income** | **351** | **86349.0** | **52082.3** | **358** | **113632.7** | **65849.8** | -6.11*** |
| Agricultural | 216 | 38391.2 | 20523.4 | 239 | 44667.4 | 24817.0 | -2.92*** |
| Non-agricultural | 160 | 32937.6 | 31692.0 | 180 | 41122.2 | 35552.7 | -2.23** |
| Labor Sale | 253 | 52632.4 | 34358.4 | 265 | 66156.6 | 49424.4 | -3.60*** |
| Business | 65 | 47876.9 | 69643.6 | 76 | 59690.8 | 87378.9 | -0.88 |
| Donation/Begging | 37 | 8594.6 | 5856.9 | 52 | 10288.5 | 8664.4 | -1.03 |
| Debt | 178 | 22747.2 | 35504.9 | 279 | 27258.1 | 39201.7 | -1.24 |
| **Total Expenditure** | **358** | **87243.2** | **39071.9** | **358** | **120553.4** | **49607.8** | -9.98*** |
| **Consumption** | **358** | **56459.7** | **20241.8** | **358** | **68749.7** | **23587.6** | -7.48*** |
| Food | 358 | 45527.9 | 17765.3 | 358 | 54932.9 | 19497.6 | -6.75*** |
| Non-Food | 358 | 10931.7 | 4486.2 | 358 | 13816.8 | 6676.7 | -6.79*** |
| **Investment** | **353** | **31219.6** | **25910.3** | **358** | **51803.6** | **34492.4** | -8.99*** |
| Education & Training | 233 | 8404.3 | 7315.0 | 289 | 10025.3 | 9127.6 | -2.20** |
| Medical | 336 | 7468.8 | 5580.2 | 338 | 8980.5 | 6278.6 | -3.30*** |
| Agricultural | 177 | 16504.5 | 10462.7 | 210 | 21633.3 | 14874.9 | -3.85*** |
| Family Business | 33 | 13666.7 | 11781.0 | 54 | 23185.2 | 14116.9 | -3.24*** |
| HH Development | 150 | 13340.0 | 14887.1 | 179 | 17933.0 | 19800.6 | -2.34** |
| Productive Asset | 37 | 10054.1 | 10477.3 | 63 | 13642.9 | 15418.5 | -1.25 |
| Durable Goods | 15 | 8600.0 | 4372.0 | 25 | 12280.0 | 11925.2 | -1.14 |
| House Repairing | 105 | 13333.3 | 16416.7 | 123 | 14150.4 | 19790.6 | -0.34 |
| Land Purchasing | 02 | 12500.0 | 8451.5 | 13 | 23307.7 | 10395.4 | -1.39 |
| Other Investment | 115 | 10256.5 | 7931.0 | 223 | 16179.4 | 12429.5 | -4.64*** |
| Savings | 32 | 6578.1 | 4649.2 | 55 | 5101.8 | 5238.2 | 1.32 |
| **Total Households** | **358** | | | **358** | | | |

**Appendix table 4.11.3: Impact of Micro-credit through Economic Performance in terms of Formal and Informal Credit Receiving Households based on Diff-in-diff Method.**

| Indicators of Economic Status | Households Economic Status (yearly) by Micro-credit Benefits | | | | | | Diff-in-diff |
|---|---|---|---|---|---|---|---|
| | Informal | | Difference | Formal | | Difference | |
| | 2016/17 | 2019/20 | | 2016/17 | 2019/20 | | |
| | (a) | (b) | $x = \dfrac{b-a}{a} \times 100$ | (c) | (d) | $y = \dfrac{d-c}{c} \times 100$ | y-x |
| **Total Income** | **87393.7** | **103145.7** | **18.02** | **86349.0** | **113632.7** | **31.60** | **13.57** |
| Agricultural | 40267.9 | 39855.6 | -1.02 | 38391.2 | 44667.4 | 16.35 | 17.37 |
| Non-agricultural | 38189.2 | 42625.0 | 11.62 | 32937.6 | 41122.2 | 24.85 | 13.23 |
| Labor Sale | 54737.4 | 68075.8 | 24.37 | 52632.4 | 66156.6 | 25.70 | 1.33 |
| Business | 43450.0 | 42416.7 | -2.38 | 47876.9 | 59690.8 | 24.68 | 27.05 |
| Donation/Begging | 2583.3 | 7142.9 | 176.50 | 8594.6 | 10288.5 | 19.71 | -156.8 |
| Debt | 17397.4 | 21263.2 | 22.22 | 22747.2 | 27258.1 | 19.83 | -2.39 |
| **Total Expenditure** | **90948.8** | **116220.5** | **27.79** | **87243.2** | **120553.4** | **38.18** | **10.39** |
| **Consumption** | **56252.0** | **66952.8** | 19.02 | **56459.7** | **68749.7** | 21.77 | 2.74 |
| Food | 45842.5 | 53409.5 | 16.51 | 45527.9 | 54932.9 | 20.66 | 4.15 |
| Non-Food | 10409.5 | 13543.3 | 30.11 | 10931.7 | 13816.8 | 26.39 | -3.71 |
| **Investment** | **35252.0** | **49267.7** | **39.76** | **31219.6** | **51803.6** | **65.93** | **26.17** |
| Education & Training | 10068.8 | 11351.6 | 12.74 | 8404.3 | 10025.3 | 19.29 | 6.55 |
| Medical | 9032.3 | 10975.8 | 21.52 | 7468.8 | 8980.5 | 20.24 | -1.28 |
| Agricultural | 14015.2 | 17108.9 | 22.07 | 16504.5 | 21633.3 | 31.08 | 9.00 |
| Family Business | 13444.4 | 19642.9 | 46.10 | 13666.7 | 23185.2 | 69.65 | 23.54 |
| HH Development | 10051.0 | 12681.2 | 26.17 | 13340.0 | 17933.0 | 34.43 | 8.26 |
| Productive Asset | 8583.3 | 11733.3 | 36.70 | 10054.1 | 13642.9 | 35.69 | -1.00 |
| Durable Goods | 9600.0 | 10000.0 | 4.17 | 8600.0 | 12280.0 | 42.79 | 38.62 |
| House Repairing | 7458.3 | 11320.0 | 51.78 | 13333.3 | 14150.4 | 6.13 | -45.65 |
| Land Purchasing | 12500.0 | 3000.0 | -76.00 | 12500.0 | 23307.7 | 86.46 | 162.5 |
| Other Investment | 8421.1 | 15153.9 | 79.95 | 10256.5 | 16179.4 | 57.75 | -22.20 |
| Savings | 3750.0 | 3166.7 | -15.55 | 6578.1 | 5101.8 | -22.44 | -6.89 |
| **Total Households** | **127** | | | **358** | | **485** | |

**Appendix table 4.11.4: Percentage Change of Households' Economic Performance due to Micro-credit Benefits based on *Diff-in-diff* Method.**

| Indicators of Economic Status | Households Economic Status (yearly) by Micro-credit Benefits | | | | | | Diff-in-diff |
| | Non-Borrower | | Difference | Borrower | | Difference | |
| | 2016/17 | 2019/20 | | 2016/17 | 2019/20 | | |
| | (a) | (b) | $x = \frac{b-a}{a} \times 100$ | (c) | (d) | $y = \frac{d-c}{c} \times 100$ | y-x |
| **Total Income** | **86885.5** | **108390.4** | **24.75** | **86626.6** | **110886.6** | **28.01** | **3.25** |
| Agricultural | 44344.8 | 46885.4 | 5.73 | 38916.7 | 43351.1 | 11.39 | 5.67 |
| Non-agricultural | 70455.7 | 70923.9 | 0.66 | 33923.9 | 41395.5 | 22.02 | 21.36 |
| Labor Sale | 45423.6 | 58521.3 | 28.83 | 53224.4 | 66678.6 | 25.28 | -3.56 |
| Business | 45960.3 | 49322.4 | 7.32 | 46835.3 | 55545.0 | 18.60 | 11.28 |
| Donation/Begging | 6800.0 | 8525.0 | 25.37 | 7755.8 | 9915.3 | 27.84 | 2.48 |
| Debt | 15031.3 | 16095.2 | 7.08 | 21117.2 | 25519.1 | 20.85 | 13.77 |
| **Total Expenditure** | **77344.3** | **97927.6** | **26.61** | **88213.5** | **119418.8** | **35.37** | **8.76** |
| **Consumption** | **50530.7** | **63399.1** | **25.47** | **56405.3** | **68279.2** | **21.05** | **-4.42** |
| Food | 41258.8 | 51561.4 | 24.97 | 45610.3 | 54534.0 | 19.57 | -5.41 |
| Non-Food | 9271.9 | 11837.7 | 27.67 | 10795.0 | 13745.2 | 27.33 | -0.34 |
| **Investment** | **27171.1** | **34528.5** | **27.08** | **32274.1** | **51139.6** | **58.45** | **31.38** |
| Education & Training | 7709.0 | 8971.3 | 16.37 | 8829.7 | 10342.9 | 17.14 | 0.76 |
| Medical | 7951.5 | 9115.8 | 14.64 | 7890.2 | 9516.0 | 20.61 | 5.96 |
| Agricultural | 15750.0 | 18261.1 | 15.94 | 15611.6 | 20164.0 | 29.16 | 13.22 |
| Family Business | 16850.0 | 24176.5 | 43.48 | 13619.1 | 22455.9 | 64.89 | 21.40 |
| HH Development | 12645.6 | 11516.3 | -8.93 | 12530.2 | 16471.8 | 31.46 | 40.39 |
| Productive Asset | 3833.3 | 4812.5 | 25.54 | 9693.9 | 13275.6 | 36.95 | 11.40 |
| Durable Goods | 6947.4 | 3000.0 | -56.82 | 9000.0 | 11500.0 | 27.78 | 84.60 |
| House Repairing | 14452.8 | 11823.1 | -18.20 | 11833.3 | 13332.4 | 12.67 | 30.86 |
| Land Purchasing | 18333.3 | 34400.0 | 87.64 | 12500.0 | 21857.1 | 74.86 | -12.8 |
| Other Investment | 7900.0 | 13370.0 | 69.24 | 9648.3 | 15947.9 | 65.29 | -3.95 |
| Savings | 26925.0 | 25181.0 | -6.48 | 6012.5 | 4911.5 | -18.31 | -11.8 |
| **Total Households** | **485** | | | **228** | | **713** | |

**Appendix table 4.13: Descriptive statistics of the Attitude of Borrowers on Micro-credits.**

| Statements | Type of Credits | | | | | | Overall | |
|---|---|---|---|---|---|---|---|---|
| | Formal | | | Informal | | | | |
| | Disagree | Neutral | Agree | Disagree | Neutral | Agree | Disagree | Agree |
| The rate of interest of micro-credit is reasonable | 74.6 | 8.4 | 17.0 | 94.5 | 2.4 | 3.1 | 79.8 | 13.4 |
| Amount of credit is sufficient | 47.5 | 17.0 | 35.5 | 77.2 | 3.1 | 19.7 | 55.3 | 31.3 |
| Duration of credit is sufficient | 61.7 | 17.0 | 21.2 | 78.0 | 3.1 | 18.9 | 66.0 | 20.6 |
| Terms and conditions are not rigid | 30.4 | 36.3 | 33.2 | 57.5 | 18.1 | 24.4 | 37.5 | 30.9 |
| By micro-finance your food security has increased | 16.8 | 38.8 | 44.4 | 53.5 | 15.0 | 31.5 | 26.4 | 41.0 |
| By micro-finance your income has increased | 27.1 | 27.7 | 45.3 | 54.3 | 19.7 | 26.0 | 34.2 | 40.2 |
| By micro-finance your savings has increased | 37.4 | 30.2 | 32.4 | 66.9 | 23.6 | 9.4 | 45.2 | 26.4 |
| Micro-finance is helping you in better access to education | 19.0 | 39.1 | 41.9 | 55.9 | 26.0 | 18.1 | 28.7 | 35.7 |
| Micro-finance is helping you in better access to healthcare | 18.7 | 36.0 | 45.3 | 52.8 | 22.8 | 24.4 | 27.6 | 39.8 |
| Micro-finance is helping you in better financial situation of your family | 17.0 | 35.2 | 47.8 | 29.1 | 16.5 | 54.3 | 20.2 | 49.5 |
| Operational assistance received from MFIs was helpful to run the business | 8.4 | 59.2 | 32.4 | 20.5 | 22.0 | 57.5 | 11.5 | 39.0 |
| Due to micro-finance, employment opportunities have been increased | 15.9 | 32.4 | 51.7 | 56.7 | 18.9 | 24.4 | 26.6 | 44.5 |
| Local loans are easier to get than MFIs | 46.4 | 15.1 | 38.5 | 21.3 | 0.0 | 78.7 | 39.8 | 49.1 |
| Local lenders are friendly than MFIs | 67.3 | 14.0 | 18.7 | 41.7 | 5.5 | 52.8 | 60.6 | 27.6 |
| Cost of local loans is lower than MFIs | 78.8 | 15.1 | 6.1 | 88.2 | 8.7 | 3.1 | 81.2 | 5.4 |
| Terms and conditions of local loans are easier than MFIs | 63.1 | 22.6 | 14.2 | 55.1 | 17.3 | 27.6 | 61.0 | 17.7 |
| **Total (n)** | **385** | | | **127** | | | **485** | |